\documentclass[12pt]{iopart}

\usepackage{iopams}
\usepackage{graphicx}
\usepackage{subfigure}
\usepackage{mcite}

\begin{document}

\title{Unconventional non-Fermi liquid state caused by nematic
criticality in cuprates}

\author{Jing-Rong Wang$^{1,2}$, Guo-Zhu Liu$^{1,4}$, and
Chang-Jin Zhang$^{2,3}$ \footnote[0]{$^4$ Author to whom any
correspondence should be addressed.}}

\address{$^1$Department of Modern Physics, University of Science and
Technology of China, Hefei, Anhui, 230026, P.R. China}

\address{$^2$High Magnetic Field Laboratory, Hefei Institutes of Physical Science,
Chinese Academy of Sciences, Hefei, 230031, China}

\address{$^3$Collaborative Innovation Center of Advanced
Microstructures, Nanjing University, Nanjing, 210093, China}

\ead{gzliu@ustc.edu.cn}

\begin{abstract}
At the nematic quantum critical point that exists in the
$d_{x^2-y^2}$-wave superconducting dome of cuprates, the massless
nodal fermions interact strongly with the quantum critical
fluctuation of nematic order. We study this problem by means of
renormalization group approach and show that, the fermion damping
rate $\left|\mathrm{Im}\Sigma^R(\omega)\right|$ vanishes more
rapidly than the energy $\omega$ and the quasiparticle residue
$Z_f\rightarrow 0$ in the limit $\omega \rightarrow 0$. The nodal
fermions thus constitute an unconventional non-Fermi liquid that
represents an even weaker violation of Fermi liquid theory than a
marginal Fermi liquid. We also investigate the interplay of quantum
nematic critical fluctuation and gauge-potential-like disorder, and
find that the effective disorder strength flows to the strong
coupling regime at low energies. Therefore, even an arbitrarily weak
disorder can drive the system to become a disorder controlled
diffusive state. Based on these theoretical results, we are able to
understand a number of interesting experimental facts observed in
curpate superconductors.
\end{abstract}


\maketitle


\section{Introduction}

A large amount of experimental and theoretical studies have been
devoted to studying the unusual properties of high temperature
cuprate superconductors in the past thirty years \cite{Orenstein00,
Lee06, Keimer15, Tsuei00, Damascelli03, Norman05, Uemura04,
Norman11, Scalapino12}. Although some consensuses have been reached,
many fundamental problems are still in debate, including the
microscopic pairing mechanism \cite{Orenstein00, Lee06, Uemura04,
Norman11, Scalapino12}, the origin of pseudogap \cite{Lee06,
Norman05}, and the description of non-Fermi liquid behaviors of the
normal state \cite{Orenstein00,Lee06,Keimer15}. In the past decade,
there have been accumulating experimental evidences for the
existence of a strong anisotropy in many of the physical properties
of YBa$_{2}$Cu$_{3}$O$_{6+\delta}$ (YBCO) \cite{Ando02, Hinkov08,
Daou10} and Bi$_{2}$Sr$_{2}$CaCu$_{2}$O$_{8+\delta}$ (BSCCO)
\cite{Lawler10, Fujita14}. Such an anisotropy is widely believed to
be driven by the formation of a novel electronic nematic order
\cite{Kivelson98, Kivelson03, Vojta09,Fradkin10,Fradkin15}, which
spontaneously breaks the $C_4$ symmetry down to a $C_2$ symmetry. In
case the nematic transition line goes across the superconducting
transition line and penetrates into the superconducting dome, there
exists a zero-temperature nematic quantum critical point (QCP). The
nematic quantum phase transition and the associated quantum critical
behaviors have been investigated extensively in recent years
\cite{Vojta00A, Vojta00B, Vojta00C, Kim08, Huh08, Xu08, Fritz09,
Wang11, Liu12,Wang2013NJP1, Wang2013NJP2, She15}.

From a theoretical perspective, there are two widely studied
scenarios to induce an electronic nematic order. First, the nematic
order can be generated by melting a stripe order that spontaneously
breaks both translational and rotational symmetry \cite{Kivelson98,
Kivelson03, Vojta09, Fradkin10, Fradkin15}. The other way is related
to Pomeranchuk instability which refers to the deformation of the
shape of the Fermi surface of a metal due to Coulomb interaction
\cite{Kivelson98, Kivelson03, Vojta09, Fradkin10, Fradkin15,
Pomeranchuk58, Yamase00A, Yamase00B, Halboth00}. In the simplest
case, Pomeranchuk instability occurs when the circular Fermi surface
of a two-dimensional metal becomes ellipse-like via quadrupolar distortion.
\cite{Kivelson98, Kivelson03, Vojta09, Fradkin10,
Fradkin15}. The Hubbard model defined on square lattices
\cite{Halboth00, Zanchi96, Maier05, Gull13, Huscroft01, Macridin06,
Gull10} provides a pertinent platform to investigate the electronic
nematic order. Halboth and Metzner \cite{Halboth00} studied a
two-dimensional Hubbard model using functional renormalization group
(RG) method, and revealed Pomeranchuk instability and nematic order.
More recently, the square lattice Hubbard model is studied by
variational cluster approximation \cite{Fang13, Fang15} and found to
display a local nematic phase under certain circumstances, which
might be applied to understand the intra-unit-cell electronic
nematicity observed in the scanning tunneling spectroscopy (STS)
measurements by Lawler \emph{et al} \cite{Lawler10}.

Another context of studying nematic order is provided by the
superconducting dome of cuprate superconductors. For a pure
$d_{x^2-y^2}$-wave superconductor, the discrete $C_{4}$ symmetry is
preserved. However, when superconductivity coexists with a nematic
order, the gap nodes are shifted from their original positions and
the $C_{4}$ symmetry is broken down to $C_{2}$ \cite{Vojta00A,
Vojta00C, Kim08, Huh08}. Such an anisotropic superconducting state
is physically equivalent to a $d_{x^2-y^2}+s$-wave superdconducting
state \cite{Vojta00A, Vojta00C}. At the nematic QCP, the massless
fermions excited from the $d_{x^2 - y^2}$-wave gap nodes couple
strongly to the quantum critical fluctuation of nematic order
parameter, which can be effectively described by a (2+1)-dimensional
field theory \cite{Vojta00A, Vojta00B, Vojta00C, Kim08, Huh08, Xu08,
Fritz09, Wang11, Liu12, Wang2013NJP1, Wang2013NJP2, She15}. This
model was first analyzed by Vojta \emph{et al.} \cite{Vojta00A,
Vojta00B, Vojta00C}, who made an $\epsilon$-expansion and argued
that the nematic phase transition is turned to first order. Kim
\emph{et al.} \cite{Kim08} later tackled the same model by means of
$1/N$-expansion, where $N$ is a large fermion flavor, and concluded
that the transition remains continuous. Huh and Sachdev \cite{Huh08}
performed a renormalization group (RG) analysis, and showed that the
fermion velocity ratio $v_{\Delta}/v_{F} \rightarrow 0$ in
the lowest energy limit, where $v_{F}$ is the Fermi velocity of
nodal fermions and $v_{\Delta}$ the gap velocity \cite{Vojta00A,
Vojta00B, Vojta00C, Kim08, Huh08, Xu08, Fritz09, Wang11, Liu12,
Wang2013NJP1, Wang2013NJP2, She15}. The unusual velocity
renormalization leads to significant changes of a number of
quantities, including density of states (DOS) \cite{Xu08}, specific
heat \cite{Xu08}, low-$T$ thermal conductivity \cite{Fritz09},
superfluid density \cite{Liu12}, and London penetration depth
\cite{She15}.

In this article, we revisit the issue of unusual physical properties
caused by the strong interaction between massless nodal fermions and
critical nematic fluctuation. We will apply the powerful RG approach
to calculate the fermion damping rate
$\left|\mathrm{Im}\Sigma^R(\omega)\right|$, where
$\mathrm{Im}\Sigma^R(\omega)$ is the imaginary part of retarded
fermion self-energy, and the corresponding quasiparticle residue
$Z_f$. Kim \emph{et al.} \cite{Kim08} have previously computed the
fermion self-energy and spectral function. Interestingly, it will
become clear below that the RG analysis give rise to different
results once the singular renormalization of fermion velocities is
taken into account. In particular, we will show that the fermion
damping rate vanishes upon approaching the Fermi level more rapidly
than the energy $\omega$, namely $\lim_{\omega \rightarrow
0}\left|\mathrm{Im} \Sigma^R(\omega)\right|/\omega \rightarrow 0$.
According to the conventional notion of quantum many-particle
physics, one would expect the system to behave as a normal Fermi
liquid. However, by analyzing the RG results, we find that the
quasiparticle residue vanishes, i.e., $Z_f \rightarrow 0$, in the
limit $\omega \rightarrow 0$. Therefore, the system is actually a
non-Fermi liquid that represents an even weaker violation of Fermi
liquid theory comparing to a marginal Fermi liquid (MFL). To the
best of our knowledge, this type of unconventional non-Fermi liquid
behavior has not been reported previously.

In realistic materials, there are always certain amount of
disorders, which may play an important role. As demonstrated earlier
by Nersesyan \emph{et al.} \cite{Nersesyan94, Nersesyan95}, the most
important disorder in cuprates behaves like a randomly distributed
gauge potential. Thus, we will mainly study the influence of random
gauge potential on the physical properties of nodal fermions and
also the stability of nematic QCP. In the absence of nematic order,
the coupling between nodal fermions and random gauge potential has
attracted considerable interest \cite{Nersesyan94, Nersesyan95,
Altland02}. Here, we consider the case in which nodal fermions
couple to both the nematic order and random gauge potential, and
then study the interplay of these two interactions by means of RG
method. We find that the effective strength of gauge potential
disorders tends to diverge at the lowest energy. This behavior
signals the emergence of a finite zero-energy DOS $\rho(0)$ and the
happening of quantum phase transition from an unconventional
non-Fermi liquid to a disorder dominated diffusive state. The nodal
fermions acquire a finite scattering rate $\gamma$, which in turn
affects the thermodynamic and spectral behaviors of nodal fermions.

The RG results for the self-energy of nodal fermions can be used to
understand a number of experimental facts observed in cuprate
superconductors. We will show that the RG  results are qualitatively consistent
with some recent measurements of specific heat, fermion damping
rate, and temperature dependence of fermion velocities.

The rest of the paper will be organized as follows. We first present
the effective low-energy field theory for the interaction between
nematic order and nodal fermions in section~\ref{Sec:ModelDerivation}.
The random gauge potential is also introduced in this section. In
section~\ref{Sec:RGAnalysisClean}, we make detailed theoretical
analysis for the self-consistent RG equations of the fermion
velocities in the clean case. Based on the RG solutions, we proceed
to compute the fermion damping rate, quasiparticle residue, and
other physical quantities. We will show that the quantum critical
fluctuation of nematic order leads to unconventional non-Fermi
liquid behaviors of nodal fermions. We consider the influence of
random gauge potential in section~\ref{Sec:RGAnalysisDisorder} and find
that the effective disorder strength flows to the strong coupling
regime at low energies. Therefore, even weak disorders play a
significant role and drive the system to enter into a disorder
controlled diffusive state. In section~\ref{Sec:CompaisionExperiments},
we discuss the possible application of the RG results to
some experimental findings of cuprates. In section~\ref{Sec:Summary},
we briefly summarize our main results.

\section{Effective field theory\label{Sec:ModelDerivation}}

We start from an effective action $S = S_{\Psi} + S_{\phi} +
S_{\Psi\phi}$. The free action for the nodal fermions is
\cite{Vojta00A,Vojta00B,Vojta00C,Kim08, Huh08}
\begin{eqnarray}
S_{\Psi} &=& \int\frac{d\omega d^{2}\mathbf{k}}{(2\pi)^{3}}
\Psi^{\dagger}_{1a}(-i\omega + v_{F}k_{x} \tau^{z} +
v_{\Delta}k_{y}\tau^x)\Psi_{1a} \nonumber \\
&+& \int\frac{d\omega d^{2}\mathbf{k}}{(2\pi)^{3}}
\Psi^{\dagger}_{2a}(-i\omega + v_{F}k_{y}\tau^{z} +
v_{\Delta}k_{x}\tau^{x})\Psi_{2a},
\end{eqnarray}
where $\tau^{(x,y,z)}$ denote the Pauli matrices. The two component
Nambu spinors are defined by
$\Psi_{1a}=(f_{1a},\epsilon_{ab}f_{3b}^{\dag})^{T}$ and
$\Psi_{2a}=(f_{2a},\epsilon_{ab}f_{4b}^{\dag})^{T}$ where
$\epsilon_{ab}=-\epsilon_{ba}$. $f_{1a}$, $f_{2a}$, $f_{3a}$ and
$f_{4a}$ represent fermions excited from the nodal points $(K,K)$,
$(-K,K)$, $(-K,-K)$, and $(K,-K)$ respectively \cite{Vojta00A,
Vojta00B, Vojta00C, Kim08, Huh08}. The repeated spin index $a$ is
summed from 1 to $N$, where $N$ is the number of fermion spin
components with the physical value being $2$. The action $S_{\phi}$
describes the nematic order parameter, which is expanded in real
space as
\begin{eqnarray}
S_{\phi} = \int d\tau
d^2\mathbf{x}\left[\frac{1}{2}(\partial_{\tau}\phi)^2 +
\frac{c^2}{2}(\nabla\phi)^2 +
\frac{r}{2}\phi^2+\frac{u_0}{24}\phi^4\right],
\end{eqnarray}
where $\tau$ is imaginary time and $c$ velocity of field $\phi$. It
is convenient to choose $c=1$. Mass parameter $r$ tunes a quantum
phase transition from $d_{x^2-y^2}$-wave superconducting state to a
state where the superconducting and nematic orders coexist, with $r
= 0$ defining the QCP. Moreover, $u_0$ is the quartic
self-interaction strength. The nematic order couples to fermions via
a Yukawa-coupling term
\begin{eqnarray}
S_{\Psi\phi} = \lambda_0 \int d\tau d^2\mathbf{x}
\phi\left(\Psi^{\dagger}_{1a}\tau^{x}\Psi_{1a} +
\Psi^{\dagger}_{2a}\tau^{x}\Psi_{2a}\right),
\end{eqnarray}
where $\lambda_{0}$ is the coupling constant. In addition to the
nematic order, the nodal fermions also couple to gauge-potential
type disorders, which is described by
\begin{eqnarray}
S_{\mathrm{dis}} = v_{\Gamma}\int d^2\mathbf{x}A(\mathbf{x})
\left(\Psi_{1a}^{\dag}\Gamma \Psi_{1a} + \Psi_{2a}^{\dag}\Gamma
\Psi_{2a}\right).\label{Eq:ActionDisorder}
\end{eqnarray}
with $\Gamma=(\tau^{z},\tau^{x})$ and
$v_{\Gamma}=(v_{\Gamma1},v_{\Gamma2})$. Here, the random gauge
potential $A(\mathbf{x})$ is assumed to be a Gaussian white noise,
defined by
\begin{eqnarray}
\left<A(\mathbf{x})\right> = 0, \quad
\left<A(\mathbf{x})A(\mathbf{x}')\right> = g\delta^2(\mathbf{x} -
\mathbf{x}'),
\end{eqnarray}
where $g$ is the impurity concentration and $v_{\Gamma}$ measures
the strength of a single impurity.

We will follow Huh and Sachdev \cite{Huh08} and perform RG analysis
by employing $1/N$-expansion. The inverse of free propagator of
$\phi$ behaves as $q^2 + r$. After including the polarization, there
will be an additional linear-in-$q$ term. The $q$-term dominates at
small $q$ over $q^2$-term, so the $q^2$-term can be neglected. Near
the nematic QCP, we keep only the mass term and rescale $\phi
\longrightarrow \phi/\lambda_0$ and $r\longrightarrow N_f r
\lambda^2_0$, leading to
\begin{eqnarray}
S &=& S_{\Psi} + \int d \tau d^2\mathbf{x}
\left\{\frac{N_fr}{2}\phi^2  + \phi\left[\Psi^{\dagger}_{1a}M_{1}
\Psi_{1a} + \Psi^{\dagger}_{2a}M_{2}\Psi_{2a}\right]\right\}.
\end{eqnarray}
The free propagators for fermions $\Psi_{1a}$ and $\Psi_{2a}$ are
written as
\begin{eqnarray}
G_{1a}^{0}(\omega,\mathbf{k}) &=& \frac{1}{-i\omega +
v_{F}k_{x}\tau^{z} + v_{\Delta}k_{y}\tau^{x}}, \\
G_{2a}^{0}(\omega,\mathbf{k}) &=& \frac{1}{-i\omega +
v_{F}k_{y}\tau^{z} + v_{\Delta}k_{x}\tau^{x}},
\end{eqnarray}
respectively. At the nematic QCP, $r=0$, the propagator for the
nematic order field $\phi$ is
\begin{eqnarray}
D(\Omega,\mathbf{q}) = \frac{1}{\Pi(\Omega,\mathbf{q})},
\end{eqnarray}
where the polarization $\Pi(\Omega,\mathbf{q})$, to the leading
order of $1/N$-expansion, is given by
\begin{eqnarray}
\Pi(\Omega,\mathbf{q}) &=& \sum_{i=1,2}\sum_{a=1}^{N}
\int\frac{d\omega}{2\pi}\frac{d^{2}\mathbf{k}}{(2\pi)^{2}}
\mathrm{Tr}\left[\tau^{x} G^{0}_{ia}(\mathbf{k},
\omega)\tau^{x} G^{0}_{ia}(\mathbf{k+q},\omega +
\Omega)\right] \nonumber \\
&=& \frac{N}{16v_{F}v_{\Delta}}
\frac{\Omega^{2}+v_{F}^{2}q_{x}^{2}}{\sqrt{\Omega^2 +
v_{F}^{2}q_{x}^{2}+v_{\Delta}^{2}q_{y}^{2}}} + (q_{x}\leftrightarrow
q_{y}).
\end{eqnarray}

After carrying out RG calculations \cite{Huh08, Wang11}, we find
that all the physical parameters flow with a varying length scale
$l$ as follows
\begin{eqnarray}
\frac{dv_{F}}{dl} &=& \left(C_{1} - C_{2} -
C_{g}\right)v_{F},\label{Eq:VRGVF}
\\
\frac{dv_{\Delta}}{dl} &=& \left(C_{1} - C_{3} -
C_{g}\right)v_{\Delta},\label{Eq:VRGVDelta}
\\
\frac{d(v_{\Delta}/v_{F})}{dl} &=&
(C_{2}-C_{3})\frac{v_{\Delta}}{v_{F}},\label{Eq:VRGVRatio}
\\
\frac{dv_{\Gamma1}}{dl} &=& \left(C_{1} - C_{2} -
C_{g}\right)v_{\Gamma1},\label{Eq:VRGVGammaGP1}
\\
\frac{dv_{\Gamma2}}{dl} &=& \left(C_{1} - C_{3} -
C_{g}\right)v_{\Gamma2},\label{Eq:VRGVGammaGP2}
\end{eqnarray}
where we have introduced a quantity
\begin{eqnarray}
C_g = \frac{v_{\Gamma1}^{2} + v_{\Gamma2}^{2}}{2\pi
v_{F}v_{\Delta}}g.
\end{eqnarray}
The expressions of $C_{1,2,3}$ can be found in the \ref{App:C1C2C3}.
We emphasize that the impact of disorders is characterized by the
quantity $C_{g}$, rather than $v_{\Gamma i}$ with $i=1,2$. It will be shown below
that $C_{g}$ flows to strong coupling at large $l$, leading to
remarkable physical properties.

\section{Unconventional non-Fermi liquid behaviors of nodal fermions
\label{Sec:RGAnalysisClean}}

It is well established that conventional metals can be described by
the standard Fermi liquid theory, which states that the fermionic
excitations of a normal Fermi liquid must have a sufficiently long
lifetime and exhibit a sharp quasiparticle peak in their spectral
peak despite the existence of Coulomb interaction \cite{Giuliani05}.
The conventional notion is that the fermionic quasiparticles
constitute a normal Fermi liquid if their zero-$T$ damping rate
$\left|\mathrm{Im}\Sigma^R(\omega)\right|$ vanishes more rapidly
than $\omega$ in the limit $\omega \rightarrow 0$. This criterion
can be mathematically expressed as
\begin{eqnarray}
\lim_{\omega \rightarrow 0}\frac{\mathrm{Im}
\Sigma^{R}(\omega)}{\omega} \rightarrow 0.
\label{Eq:DampingTradi}
\end{eqnarray}
Another criterion to identify normal Fermi liquid is to define an
important quantity: the quasiparticle residue, also called
renormalization factor, $Z_f$. The residue $Z_f$ is usually
calculated through the definition
\begin{eqnarray}
Z_{f}(\omega) = \frac{1}{\left|1 - \frac{\partial}{\partial\omega}
\mathrm{Re}\Sigma^{R}(\omega)\right|},\label{Eq:ZDefA}
\end{eqnarray}
where the real part of retarded fermion self-energy
$\mathrm{Re}\Sigma^{R}(\omega)$ is related to
$\mathrm{Im}\Sigma^{R}(\omega)$ via the Kramers-Kronig (K.-K.)
relationship. The residue $Z_f$ is finite in a normal Fermi liquid,
but vanishes in a non-Fermi liquid. Generically, the fermion damping
rate in an interacting fermion system can be formally written as
\begin{eqnarray}
\mathrm{Im}\Sigma^{R}(\omega) = C_F |\omega|^x,
\end{eqnarray}
where $C_F$ is a constant. With the help of K.-K. relation, the
corresponding real part of the fermion self-energy, in the low
energy limit, is given by
\begin{eqnarray}
\mathrm{Re}\Sigma^{R}(\omega)=\left\{\begin{array}{lll}
C_F\mathrm{sgn}(\omega)|\omega|^{x}I(x), &\mathrm{if}& 0 < x < 1, \\
C_F\frac{2}{\pi}\omega\ln\left(\frac{\omega_0}{|\omega|}\right), &
\mathrm{if} & x=1 ,\\
C_F\frac{2}{\pi}\frac{|\omega_0|^{x-1}}{x-1}\omega, & \mathrm{if} &
x > 1,
\end{array}\right.
\end{eqnarray}
where $\omega_{0}$ is a cutoff, and $I(x)$ is a function that
depends only on $x$.

It can be easily checked that $Z_f = 0$ for $0 < x \leq 1$ and $Z_f
\neq 0$ for $x > 1$. Therefore, the above two criteria are actually
equivalent because $Z_f$ automatically takes a finite value whenever
equation (\ref{Eq:DampingTradi}) is fulfilled.

\begin{figure}[htbp]
\center
\includegraphics[width=2.8in]{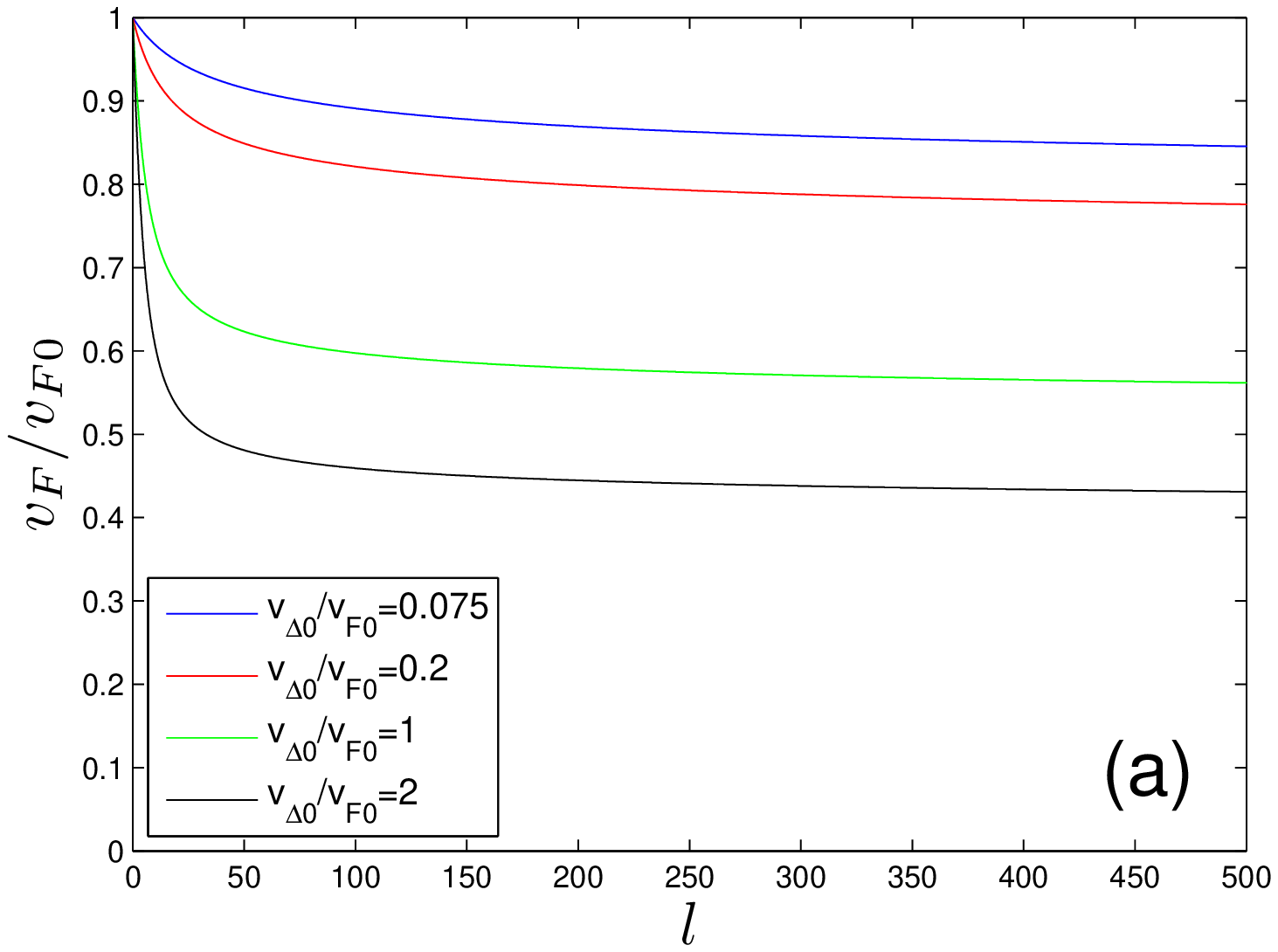}
\includegraphics[width=2.8in]{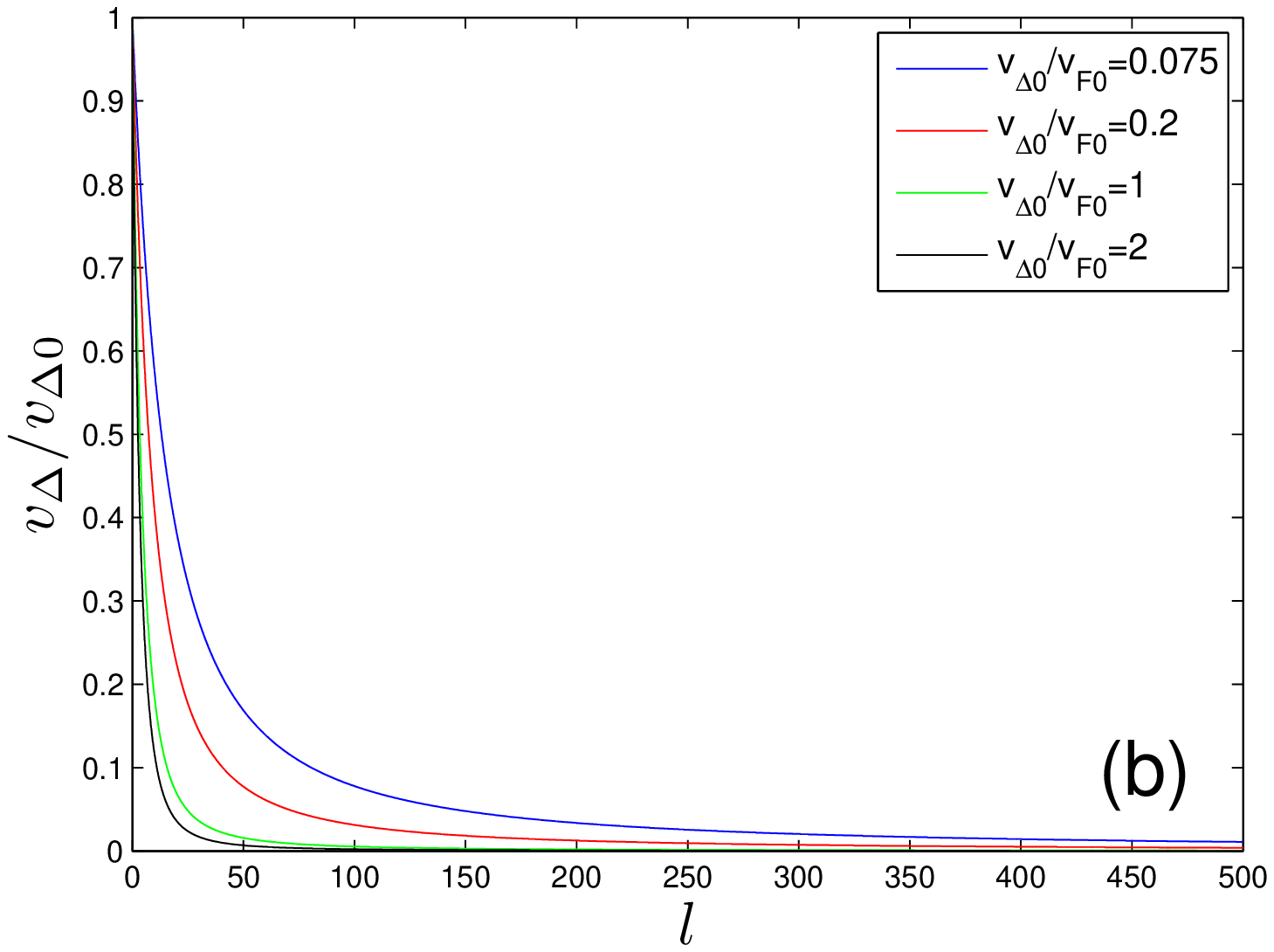}
\includegraphics[width=2.8in]{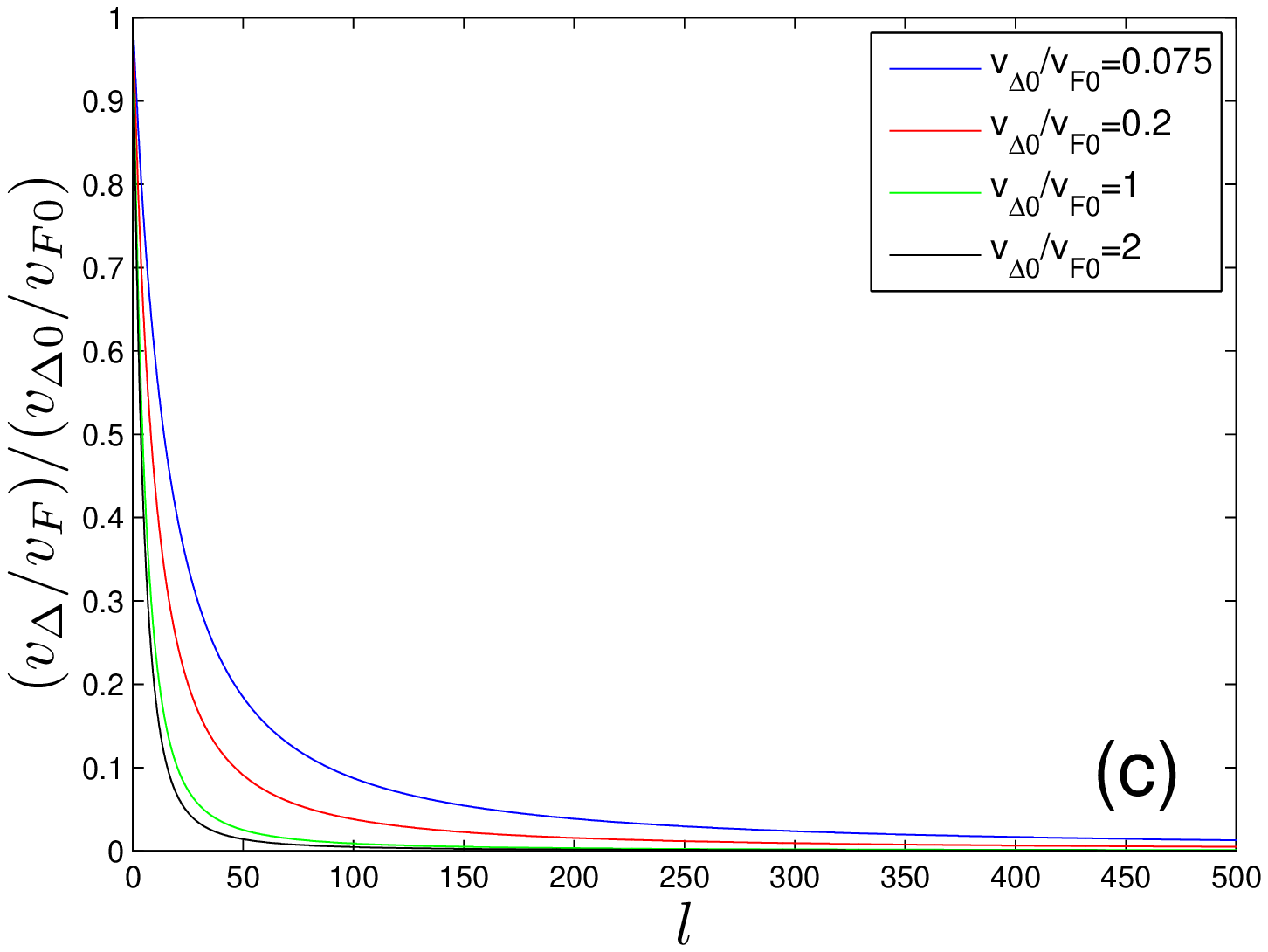}
\caption{(a), (b), and (c) show how $v_{F}$, $v_{\Delta}$, and
$v_{\Delta}/v_{F}$ flow with a varying length scale $l$ respectively
in the clean case. The initial values of velocity ratio are
$v_{\Delta0}/v_{F0} = 0.075, 0.2, 1, 2$.\label{Fig:VRGVelocity}}
\end{figure}

However, in the present nodal fermion system, the above two criteria
are no longer equivalent. We will show by explicit calculations in
this section that the damping rate of nodal fermions vanishes more
rapidly than the energy, but the residue $Z_f$ vanishes, i.e., $Z_f
\rightarrow 0$.

\subsection{Fermion damping rate and quasiparticle residue}

Now we calculate the fermion damping rate and the residue $Z_f$
utilizing the solutions of the RG equations
(\ref{Eq:VRGVF})-(\ref{Eq:VRGVGammaGP2}). The unusual
renormalization of fermion velocities need to be taken into account
in an appropriate manner. We only present the results obtained in
the clean limit $C_{g} = 0$ in this section, and include the effect
of random gauge potential in the next section.

After solving RG equations (\ref{Eq:VRGVF})-(\ref{Eq:VRGVRatio})
self-consistently, we show the $l$-dependence of fermion velocities
$v_{F}$, $v_{\Delta}$ and ratio $v_{\Delta}/v_{F}$ obtained
at different initial values of ratio $v_{\Delta0}/v_{F0}$
in figures~\ref{Fig:VRGVelocity}(a), (b) and (c) respectively. All
the quantities $v_{F}$, $v_{\Delta}$, and $v_{\Delta}/v_{F}$
decrease with growing $l$ and flow eventually to zero as $l
\rightarrow +\infty$, but apparently $v_{F}$ decreases much more
slowly than $v_{\Delta}$. If we use the ratio $v_{\Delta}/v_{F}$ to
characterize the velocity anisotropy, it is clear that the nematic
order drives an extreme velocity anisotropy $v_{\Delta}/v_{F}
\rightarrow 0$. For later use, it is helpful to extract an
approximate analytical expressions for the velocities. Considering
the leading and sub-leading terms, the solution for
$v_{\Delta}/v_{F}$ is given by
\begin{eqnarray}
\frac{v_{\Delta}}{v_{F}} \sim \frac{\pi^2N}{8}\frac{1}{l\ln(c_{s}l)}=
\frac{\pi^2N}{8}\frac{1}{l\left[\ln(l)+\ln\left(c_{s}\right)\right]},
\end{eqnarray}
with $c_{s} \approx 0.3809/N$, which is consistent with the
expression in reference~\cite{Huh08}. In the long wavelength limit,
$l\rightarrow \infty$, $\ln(c_{s})/\ln(l)\rightarrow 0$, so the
sub-leading term can be ignored. Retaining only the leading term,
the asymptotic behavior of velocity ratio can be well approximated
by
\begin{eqnarray}
\frac{v_{\Delta}}{v_{F}} \sim \frac{\pi^2N}{8}\frac{1}{l\ln(l)}.
\label{Eq:VRatioApproximate}
\end{eqnarray}
Substituting the asymptotic form of $v_{\Delta}/v_{F}$
into equation~(\ref{Eq:VRGVF}), we obtain
\begin{eqnarray}
\frac{dv_{F}}{dl} \sim - \frac{\pi^2 c_{1}}{8}\frac{v_{F}}{l\ln(l)}
\end{eqnarray}
with $c_{1} \approx 0.078$ for large $l$. Solving this equation, we
express the renormalized $v_{F}$ as
\begin{eqnarray}
v_{F}(l) \sim \frac{1}{(\ln l)^{c_{2}}}\label{Eq:VFAsoBehavior}
\end{eqnarray}
with $c_{2} =\frac{\pi^2 c_{1}}{8}$. It can also be found that
$v_{\Delta}$ at large $l$ behaves approximately as
\begin{eqnarray}
v_{\Delta}(l) \sim \frac{\pi^2N}{8}\frac{1}{l\left[\ln(l)\right]^{1
+ c_{2}}}.\label{Eq:VDeltaAsoBehavior}
\end{eqnarray}
According to equations~(\ref{Eq:VFAsoBehavior}) and
(\ref{Eq:VDeltaAsoBehavior}), both $v_{F}$ and $v_{\Delta}$ vanish
as $l \rightarrow +\infty$.

To examine the impact of singular fermion velocity renormalization
and extreme velocity anisotropy on the properties of nodal fermions,
we will compute a number of physical quantities. Coming first is the
quasiparticle residue $Z_f$. Apart from the widely used definition
given by equation~(\ref{Eq:ZDefA}), the residue $Z_f$ can also be
calculated within the RG framework. The interaction induced
renormalization of fermion field $\Psi$ is encoded in the residue
$Z_f$, which exhibits the following $l$-dependence
\begin{eqnarray}
Z_{f} = e^{\int_{0}^{l}\left(C_{1}-C_{g}\right)dl'}
\end{eqnarray}
or alternatively
\begin{eqnarray}
\frac{dZ_{f}}{dl} = \left(C_{1}-C_{g}\right)Z_{f}.\label{Eq:ZFlowDef2}
\end{eqnarray}

\begin{figure}[htbp]
\center
\includegraphics[width=2.8in]{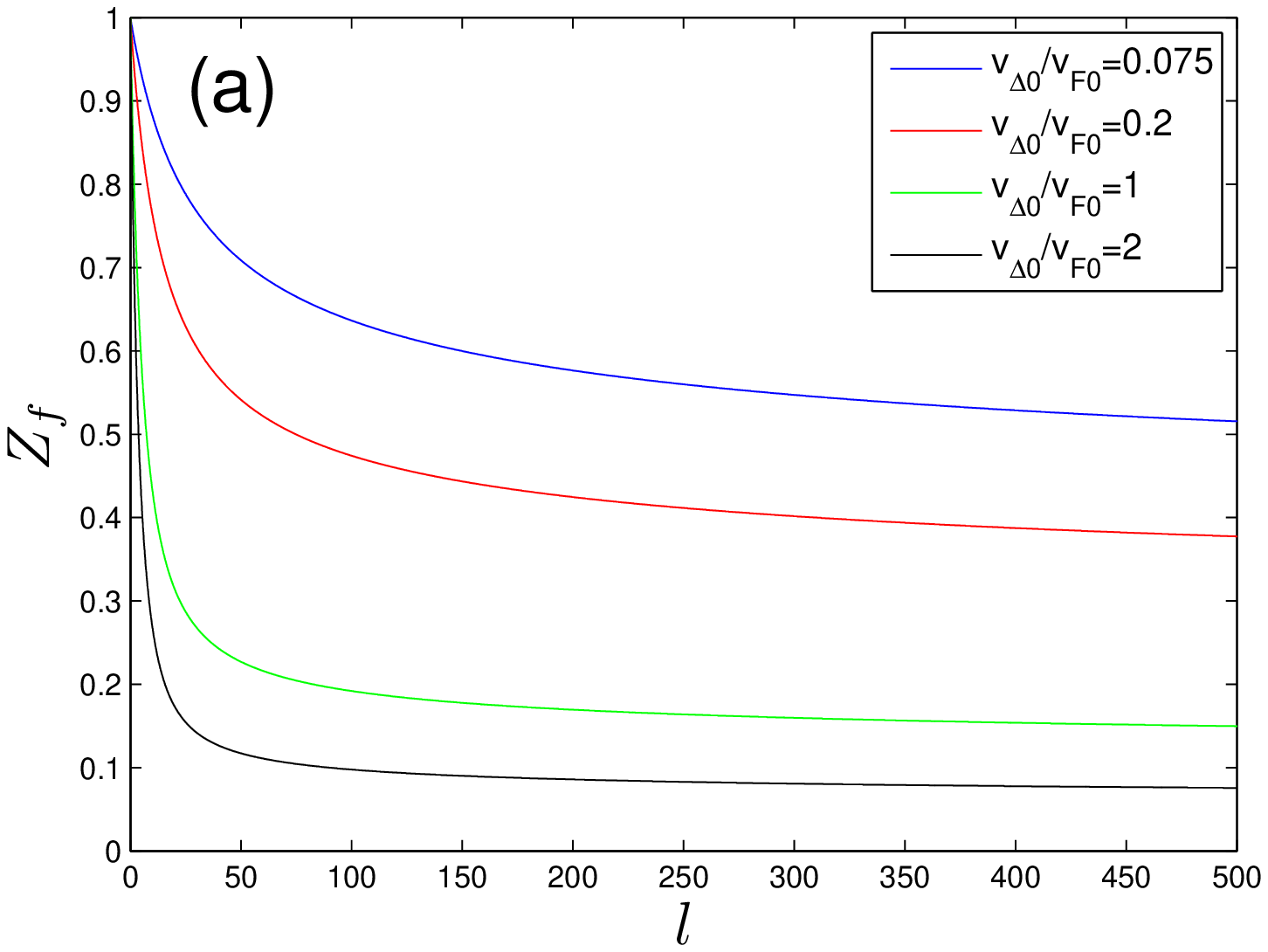}
\includegraphics[width=2.8in]{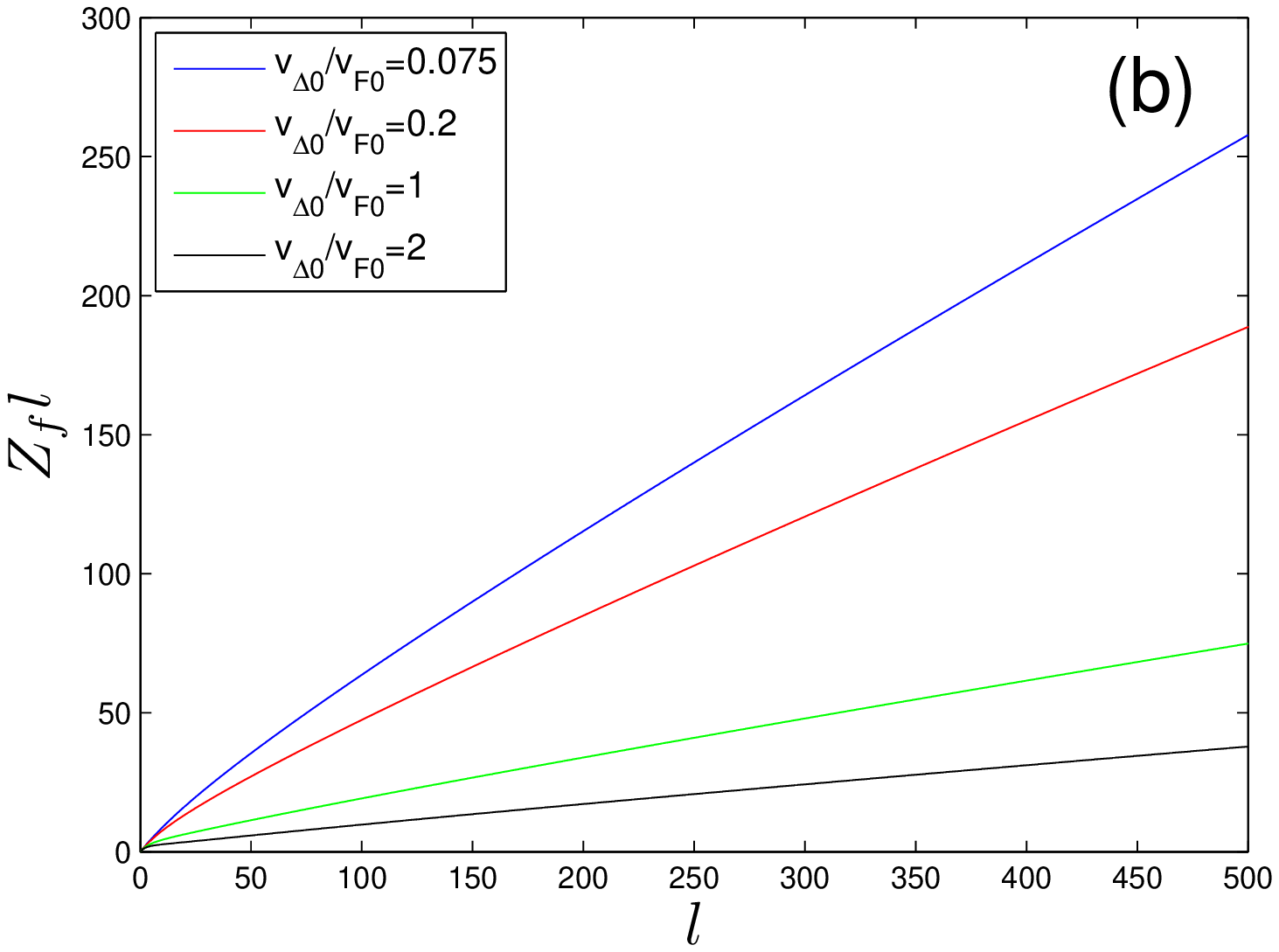}
\caption{$l$-dependence of $Z_f$ in the clean case is presented in
(a), and of $Z_{f}l$ in (b). The initial values of velocity ratio
are $v_{\Delta0}/v_{F0} = 0.075, 0.2, 1, 2$. \label{Fig:VRGZf}}
\end{figure}

The $l$-dependence of $Z_{f}$ can be easily obtained from the above
equation, and is presented in figure~\ref{Fig:VRGZf}(a). We find that
$Z_{f}$ flows to zero in the limit $l \rightarrow +\infty$, which
indicates the breakdown of normal Fermi liquid and the absence of
well-defined Landau quasiparticles. However, $Z_{f}$ decreases very
slowly with growing $l$, thus the deviation of the system from a
normal Fermi liquid ought to be quite weak. To see this point, we
plot the $l$-dependence of $Z_{f}l$ in figure~\ref{Fig:VRGZf}(b), and
find that
\begin{eqnarray}
\lim_{l\rightarrow +\infty} Z_{f}l \rightarrow +\infty.
\end{eqnarray}
Therefore, the residue $Z_{f}$ obtained at nematic QCP vanishes with
growing $l$ more slowly than that of a MFL \cite{Varma}, where
$Z_{f} \sim 1/l$. According to equation~(\ref{Eq:ZDefA}), one can
speculate that
\begin{eqnarray}
\lim_{\omega \rightarrow 0} \frac{\mathrm{Re}
\Sigma_{\mathrm{nem}}^{R}(\omega)}{\omega \ln(\omega_{0}/\omega)}
\rightarrow 0,
\end{eqnarray}
since a large length scale $l$ corresponds to a low energy $\omega$.
This speculation in turn implies that the imaginary part of the
retarded fermion self-energy
$\mathrm{Im}\Sigma_{\mathrm{nem}}^{R}(\omega)$ should display the
following low-energy behavior,
\begin{eqnarray}
\lim_{\omega \rightarrow 0}
\frac{\mathrm{Im}\Sigma_{\mathrm{nem}}^{R}(\omega)}{\omega}
\rightarrow 0.
\end{eqnarray}

To make the above discussions more quantitative, we are now going to
make a detailed analysis of the asymptotic behavior of $Z_{f}$. At
large running scale $l$, the equation of $Z_{f}$ can be approximated
by
\begin{eqnarray}
\frac{dZ_{f}}{dl} \sim -\frac{c_{4}}{N}\frac{v_{\Delta}}{v_{F}}Z_{f}
\sim -\frac{c_{5}}{l\ln l}Z_{f},
\end{eqnarray}
where $c_{4} \approx 0.426$ and $c_{5} = \frac{\pi^2 c_{4}}{8}\approx0.523$.
Solving this equation leads to the asymptotic behavior
\begin{eqnarray}
Z_{f} \sim \frac{1}{\left(\ln l\right)^{c_{5}}} \rightarrow 0
\end{eqnarray}
in the limit $l \rightarrow +\infty$. To proceed, it proves
convenient to utilize the relationship between $\omega$ and $l$:
\begin{eqnarray}
\omega = \omega_{\mathrm{0}}e^{-l},
\end{eqnarray}
where $\omega_{\mathrm{0}}$ is a cutoff. Then the real part of
retarded self-energy is approximated by
\begin{eqnarray}
\mathrm{Re}\Sigma_{\mathrm{nem}}^{R}(\omega) &\sim&
\omega\left\{\ln\left[\ln\left(\frac{\omega_{0}}{\omega}
\right)\right]\right\}^{c_{5}}\label{Eq:RealPartSelfEnergyNematicQCP}
\end{eqnarray}
as $\omega \rightarrow 0$. Using the K.-K. relation, we obtain the
imaginary part of self-energy
\begin{eqnarray}
\mathrm{Im}\Sigma_{\mathrm{nem}}^{R}(\omega) &\sim&
\frac{\pi}{2}\frac{|\omega|}{\ln\left(\frac{\omega_{0}}{\omega}\right)
\left\{\ln\left[\ln\left(\frac{\omega_{0}}{\omega}
\right)\right]\right\}^{1-c_{5}}}. \label{Eq:DampingNematicQCPCleanAppro}
\end{eqnarray}
From equation (\ref{Eq:DampingNematicQCPCleanAppro}), it is easy to
verify that this fermion damping rate is smaller than that in a MFL
and manifests the asymptotic behavior $\lim_{\omega \rightarrow 0}
\mathrm{Im}\Sigma_{\mathrm{nem}}^{R}(\omega)/\omega \rightarrow 0$,
confirming the above analysis based on numerical results. According
to the conventional notion of quantum many-body physics, one would
expect the system to behave like a normal Fermi liquid. However, in
the low energy limit $\omega\rightarrow0$, the residue $Z_{f}$
actually vanishes:
\begin{eqnarray}
Z_{f}\sim\frac{1}{\left\{\ln\left[\ln\left(\frac{\omega_{0}}{\omega}
\right)\right]\right\}^{c_{5}}}\rightarrow 0,
\label{Eq:ZNematicQCPCleanAppro}
\end{eqnarray}
which clearly implies that the system under consideration is
actually a non-Fermi liquid and the nodal fermions do not have a
well-defined quasiparticle peak in the spectral function. We see
that the above residue $Z_{f}$ flows to zero at a lower speed than
that of a MFL, i.e., $Z_{f} \sim
\frac{1}{\ln(\omega_{0}/\omega)}$. Therefore, the
quantum critical fluctuation of nematic order gives rise to an even
weaker violation of ordinary Fermi liquid theory than a MFL. To the
best of our knowledge, this sort of unconventional non-Fermi liquid
state has not been reported previously.

It is now interesting to compare this unconventional non-Fermi
liquid state with graphene. In graphene, early perturbative
calculations revealed that Dirac fermions exhibit MFL behavior with
damping rate $\mathrm{Im}\Sigma^{R}(\omega) \sim \omega$ and
vanishing $Z_f$ \cite{Gonzalez99}. Nevertheless, subsequent careful
RG studies \cite{Gonzalez99, Kotov12, WangLiu14, Hofmann14} have
found that, the fermion damping rate actually depends on energy as
$\mathrm{Im}\Sigma^{R}(\omega) \propto \omega/\ln^2\omega$ at zero
temperature due to the long-range Coulomb interaction, whereas the
corresponding $Z_{f}$ flows to a finite value as $\omega \rightarrow
0$. Therefore, graphene is a normal Fermi liquid. In contrast, the
massless nodal fermions constitute a non-Fermi liquid at the nematic
QCP, because $Z_{f}$ vanishes in the lowest energy limit. The
crucial difference between these two cases can be understood as
follows. At nematic QCP, the fermion velocities are driven to vanish
by the nematic order, so the effective fermion-nematic interaction
is significantly pronounced. In graphene, however, the fermion
velocity is dramatically enhanced at low energies by Coulomb
interaction, which then weakens the effective Coulomb interaction
and guarantees the validity of the Fermi liquid description. These
two interacting Dirac fermion systems provide interesting new
insight on the effects of strong electron correlations and also on
the criterion of non-Fermi liquid states.

Moreover, an important lesson one can learn from the research
experience of graphene is that, RG may lead to qualitatively
different spectral properties of fermions from that obtained by
ordinary perturbative expansion approach. Indeed, this is the main
motivation that has promoted us to make an extensive RG analysis of
the spectral properties of nodal fermions at the nematic QCP.

\begin{figure}[htbp]
\center
\includegraphics[width=2.85in]{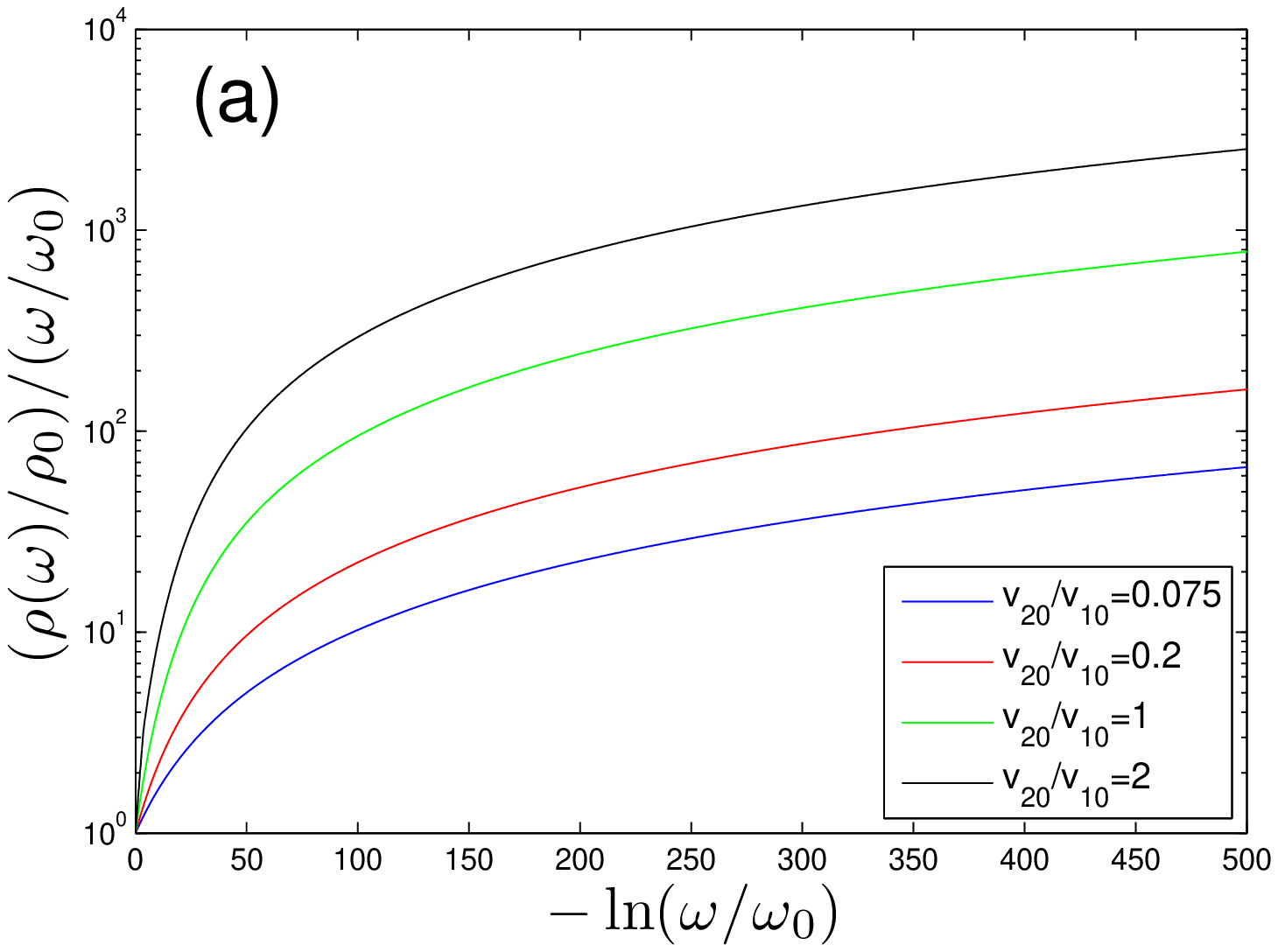}
\includegraphics[width=2.8in]{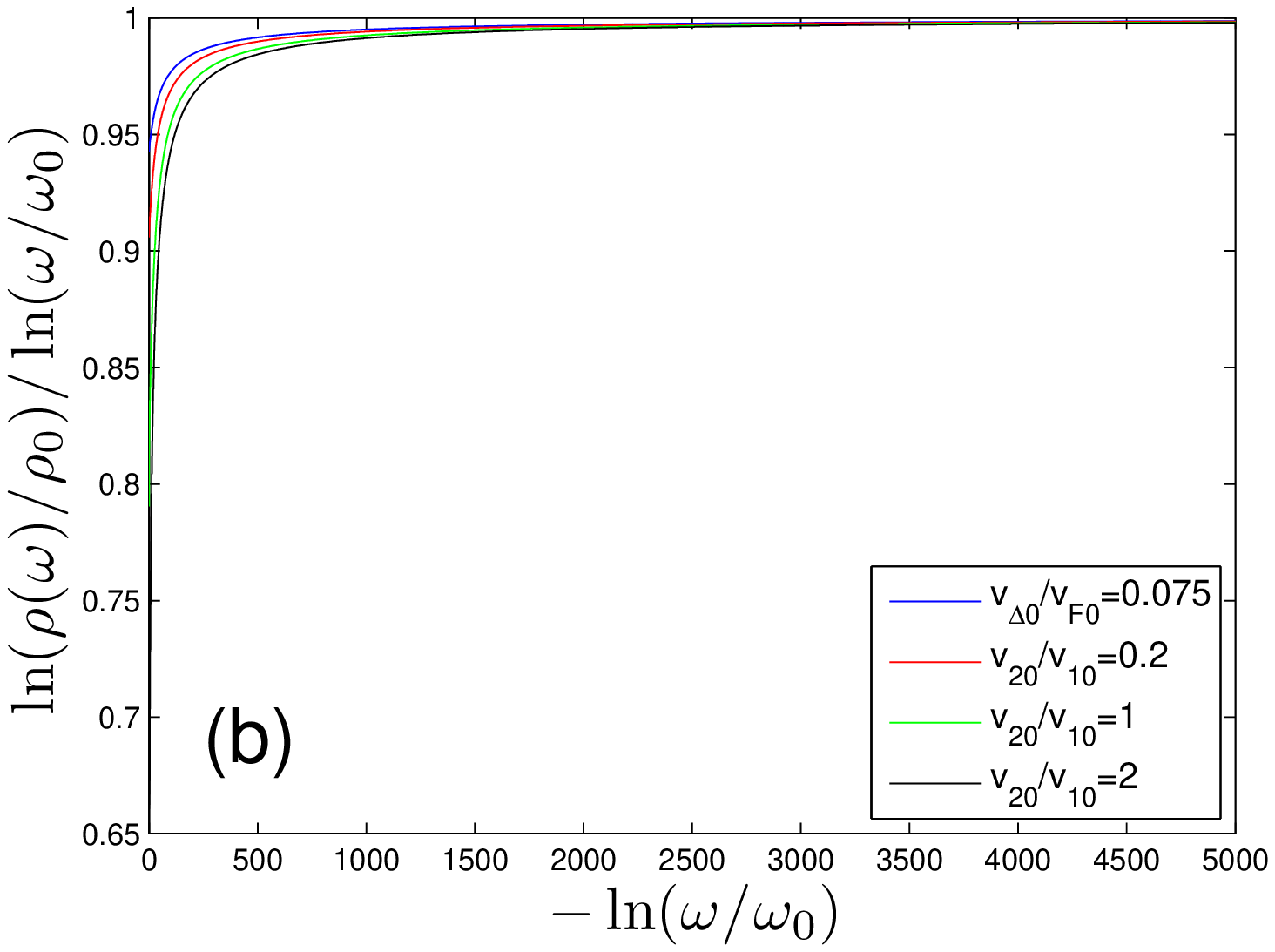}
\caption{Fermion DOS $\rho(\omega)$ in the clean limit at different
initial values $v_{\Delta0}/v_{F0}=0.075, 0.2, 1, 2.$
\label{Fig:RhoCl}}
\end{figure}

\subsection{Density of states and specific heat}

We have showed in the last subsection that the nodal fermions
exhibit unconventional non-Fermi liquid behaviors at the nematic
QCP. In this subsection, we will compute two important quantities,
namely DOS and specific heat, on the basis of the RG solutions with
the goal to gain a better understanding of the non-Fermi liquid
state.

\begin{figure}[htbp]
\center
\includegraphics[width=2.85in]{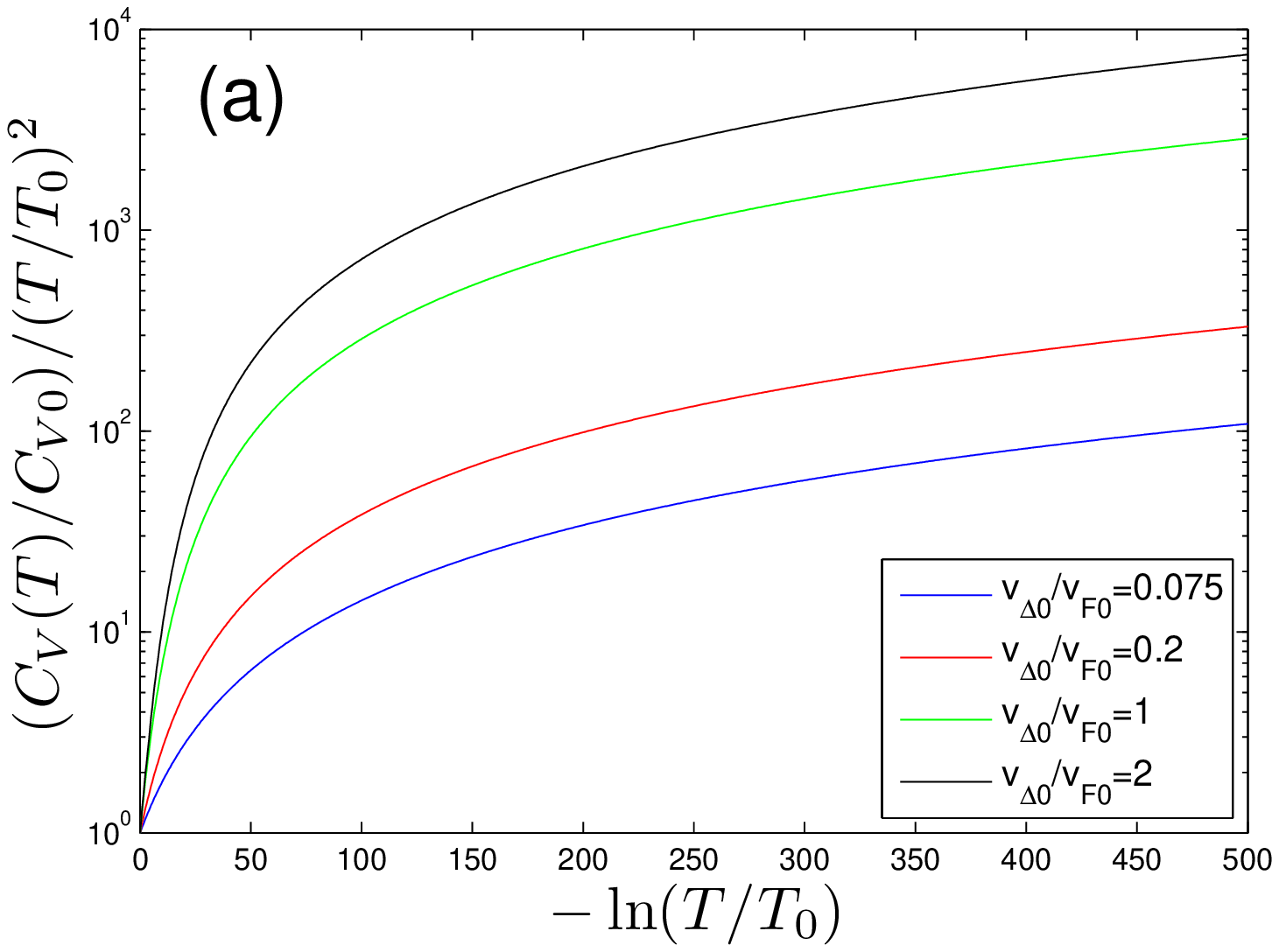}
\includegraphics[width=2.8in]{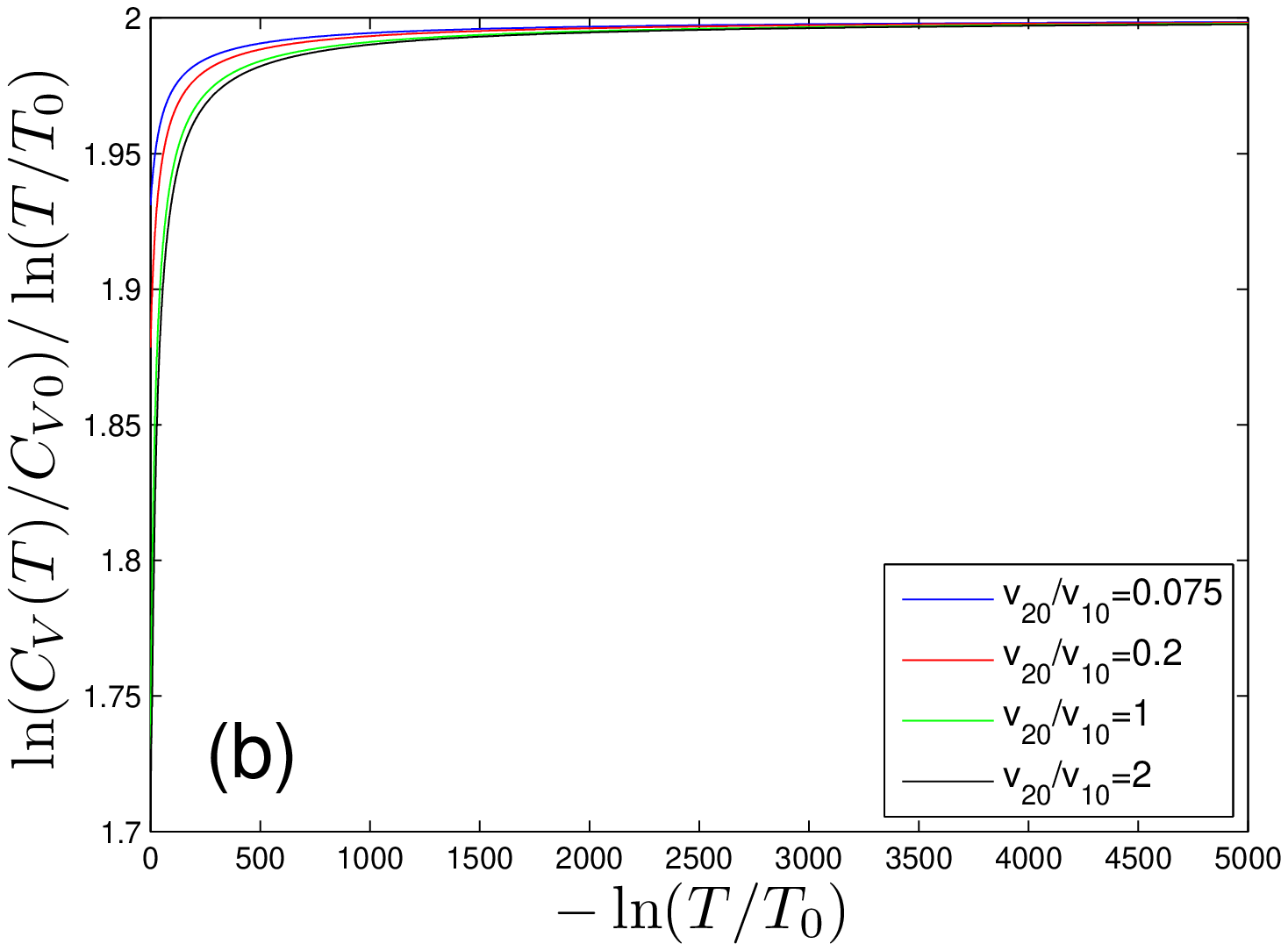}
\caption{Specific heat $C_{V}(T)$ in the clean limit at different
initial values $v_{\Delta0}/v_{F0}=0.075, 0.2, 1, 2.$
\label{Fig:CvCl}}
\end{figure}

The fermion DOS can calculated from the retarded Green functions of
nodal fermions via the definition \cite{Xu08}
\begin{eqnarray}
\fl \rho(\omega)&=&\sum_{a=1}^{N}\int\frac{dk_{x}dk_{y}}{(2\pi)^{2}}
\frac{1}{\pi}\mathrm{Tr}\left[\mathrm{Im}G_{1a}^{R}
(v_{F}k_{x},v_{\Delta}k_{y},\omega) +
\mathrm{Im}G_{2a}^{R}(v_{F}k_{x},v_{\Delta}k_{y},\omega) \right]
\nonumber \\
\fl &=& \frac{1}{v_{\Delta}v_{F}}\sum_{a=1}^{N}
\int\frac{dk_{x}'dk_{y}'}{(2\pi)^{2}}
\frac{1}{\pi}\mathrm{Tr}\left[\mathrm{Im}G_{1a}^{R}(k_{x}',k_{y}',\omega)+
\mathrm{Im}G_{2a}^{R}(k_{x}',k_{y}',\omega) \right].
\end{eqnarray}
In the absence of interactions, the fermion DOS is well-known to be
linear in $\omega$, namely $\rho(\omega)\propto \omega$. This linear
behavior will be changed once the interaction effects are
considered. After including the fermion velocity renormalization,
employing the method presented in \cite{Xu08}, the flow equation for
$\rho(\omega)$ is given by
\begin{eqnarray}
\frac{d\ln\rho}{d\ln\omega} &=&
\left\{\begin{array}{ll}
\frac{1 + C_{1} - C_{2} - C_{3} -
C_{g}}{1 - C_{1} + C_{2} + C_{g}}, & \mathrm{if}\quad v_{\Delta}<v_{F},
\\
\\
\frac{1 + C_{1} - C_{2} - C_{3} - C_{g}}{1 - C_{1} + C_{3} + C_{g}},
& \mathrm{if}\quad v_{\Delta} > v_{F}.
\end{array}\right.
\label{Eq:RhoRGEq}
\end{eqnarray}
In cuprates, the initial value of velocity ratio
$v_{\Delta0}/v_{F0}$ is known to be much smaller than $1$
\cite{Chiao00}. Additionally, due to the quantum fluctuation of
nematic order, $v_{\Delta}/v_{F}$ decreases monotonously as the
energy scale is lowering. Therefore, for a given initial value
$v_{\Delta0}/v_{F0} < 1$, we only need to consider the case
$v_{\Delta} < v_{F}$. In order to show that the conclusion is
independent of the condition $v_{\Delta0}/v_{F0} < 1$, we also plot
the curves for the case $v_{\Delta0}/v_{F0} = 2$ in figures \ref{Fig:RhoCl} and \ref{Fig:CvCl}.
The RG equations of DOS and specific heat are also given with a
generalized form. In the clean limit, we plot the results for
$\rho(\omega)$ in figure~\ref{Fig:RhoCl}. We see from
figure~\ref{Fig:RhoCl}(a) that $\rho(\omega)$ is apparently not
linear in $\omega$, but displays the asymptotic behavior
\begin{eqnarray}
\lim_{\omega \rightarrow 0} \frac{\rho(\omega)}{\omega} \rightarrow
+\infty.
\end{eqnarray}
On the other hand, figure \ref{Fig:RhoCl}(b) implies that
\begin{eqnarray}
\lim_{\omega \rightarrow 0}
\frac{\ln(\rho(\omega)/\rho_{0})}{\ln(\omega/\omega_{0})}
\rightarrow 1.
\end{eqnarray}
This DOS $\rho(\omega)$ is qualitatively different from the power
law function $\rho(\omega)\sim\omega^{1-\alpha}$, where $\alpha$ is
a small finite value, obtained previously in reference \cite{Xu08}.

In the non-interacting limit, the fermion specific heat depends on
$T$ as $C_{V}(T)\propto T^2$. As shown in \ref{App:SpecificHeat},
including the influence of the quantum fluctuation of nematic order
and random gauge potential leads us to
\begin{eqnarray}
\frac{d \ln C_{V}}{d\ln T} &=& \left\{\begin{array}{ll} 2 +
\frac{2C_{1} - C_{2} - C_{3} - 2C_{g}}{1 - C_{1} + C_{2} + C_{g}}, &
\mathrm{if}\quad v_{\Delta}<v_{F},
\\
\\
 2 + \frac{2C_{1} - C_{2} - C_{3} -
2C_{g}}{1 - C_{1} + C_{3} + C_{g}}, & \mathrm{if}\quad v_{\Delta}>v_{F}.
\end{array}\right.\label{Eq:CvRGEq}
\end{eqnarray}
At low $T$, $C_{V}(T)$ behaves as
\begin{eqnarray}
\lim_{T \rightarrow 0} \frac{C_{V}(T)}{T^{2}} \rightarrow +\infty,
\end{eqnarray}
which is visualized in figure \ref{Fig:CvCl}(a). Apparently, the
original quadratic $T$-dependence of $C_{V}(T)$ obtained in the
non-interacting limit is significantly altered. According to figure
\ref{Fig:CvCl}(b), we can express $C_{V}(T)$ at very small $T$ as
\begin{eqnarray}
\lim_{T \rightarrow 0}\frac{\ln(C_{V}(T)/C_{V0})}{\ln(T/T_{0})}
\rightarrow 2,
\end{eqnarray}
which is also distinct from the power law $T$-dependence
$C_{V}(T)\sim T^{2-\beta}$, where $\beta$ is a small finite
constant, obtained previously \cite{Xu08}.

Simple analysis reveal that the three parameters $C_{1}$, $C_{2}$,
and $C_{3}$ appearing in the RG equations (\ref{Eq:RhoRGEq}) and
(\ref{Eq:CvRGEq}) all flow to zero in the lowest energy limit
\cite{Huh08}. This asymptotic behavior makes it impossible to
express $\rho(\omega)$ and $C_V(T)$ by power law functions. As given
in \ref{App:ApproDOSCv}, the low energy behavior of $\rho(\omega)$
and the low-$T$ behavior of $C_V(T)$ can be approximately expressed
as
\begin{eqnarray}
\fl \rho(\omega)&\sim&\left(\frac{\omega}{\omega_{0}}\right)
\ln\left(\frac{\omega_{0}}{\omega}\right)
\left(\ln\ln\left(\frac{\omega_{0}}{\omega}\right)\right)^{a_{\rho}}\exp
\left[\frac{1}{2}\left(\ln\ln\ln
\left(\frac{\omega_{0}}{\omega}\right)\right)^{2}\right],
\\
\fl C_{V}(T) &\sim& \left(\frac{T}{T_{0}}\right)^{2}
\ln\left(\frac{T_{0}}{T}\right)
\left(\ln\ln\left(\frac{T_{0}}{T}\right)\right)^{a_{C}} \exp
\left[\frac{1}{2}\left(\ln\ln\ln
\left(\frac{T_{0}}{T}\right)\right)^{2}\right],
\end{eqnarray}
where $a_{\rho}$ and $a_{C}$ are two negative constants. At $N=2$,
$a_{\rho} \approx -1.896$ and $a_{C} \approx -1.466$.

We notice that our RG equation for $C_V(T)$ (\ref{Eq:CvRGEq}) is not
exactly the same as that obtained by Xu \emph{et al.} \cite{Xu08}.
This difference does not affect our conclusion that $C_V(T)$ is not
a power law function at low $T$. Indeed, if we start from the RG
equation of $C_V(T)$ presented in \cite{Xu08}, we reach the same
conclusion. A more detailed discussion is presented in
\ref{App:SpecificHeat} and \ref{App:ApproDOSCv}.

\section{Impact of Random gauge potential\label{Sec:RGAnalysisDisorder}}

In this section, we investigate the impact of random gauge potential
on the quantum critical behaviors near the nematic QCP. We will
first show the RG solutions obtained in the presence of random gauge
potential, and then re-calculate the fermion DOS and specific heat
after considering the influence of disorder scattering. To this end,
we need to retain a nonzero $C_g$ in the RG equations
(\ref{Eq:VRGVF})-(\ref{Eq:VRGVGammaGP2}), (\ref{Eq:ZFlowDef2}),
(\ref{Eq:RhoRGEq}), and (\ref{Eq:CvRGEq}), and then solve these RG
equations numerically.

\begin{figure*}[htbp]
\center
\includegraphics[width=2.7in]{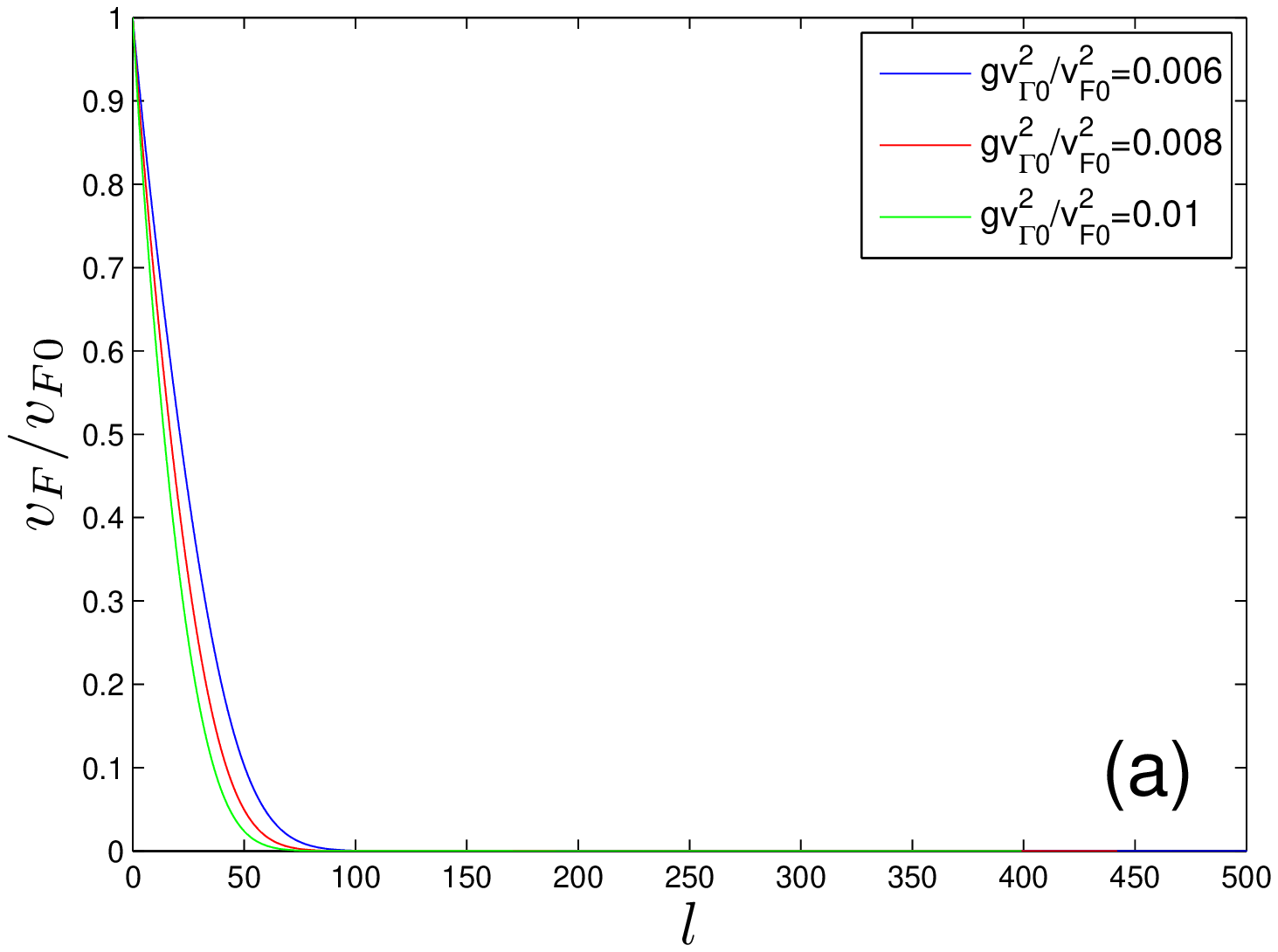}
\hspace{1.5cm}
\includegraphics[width=2.7in]{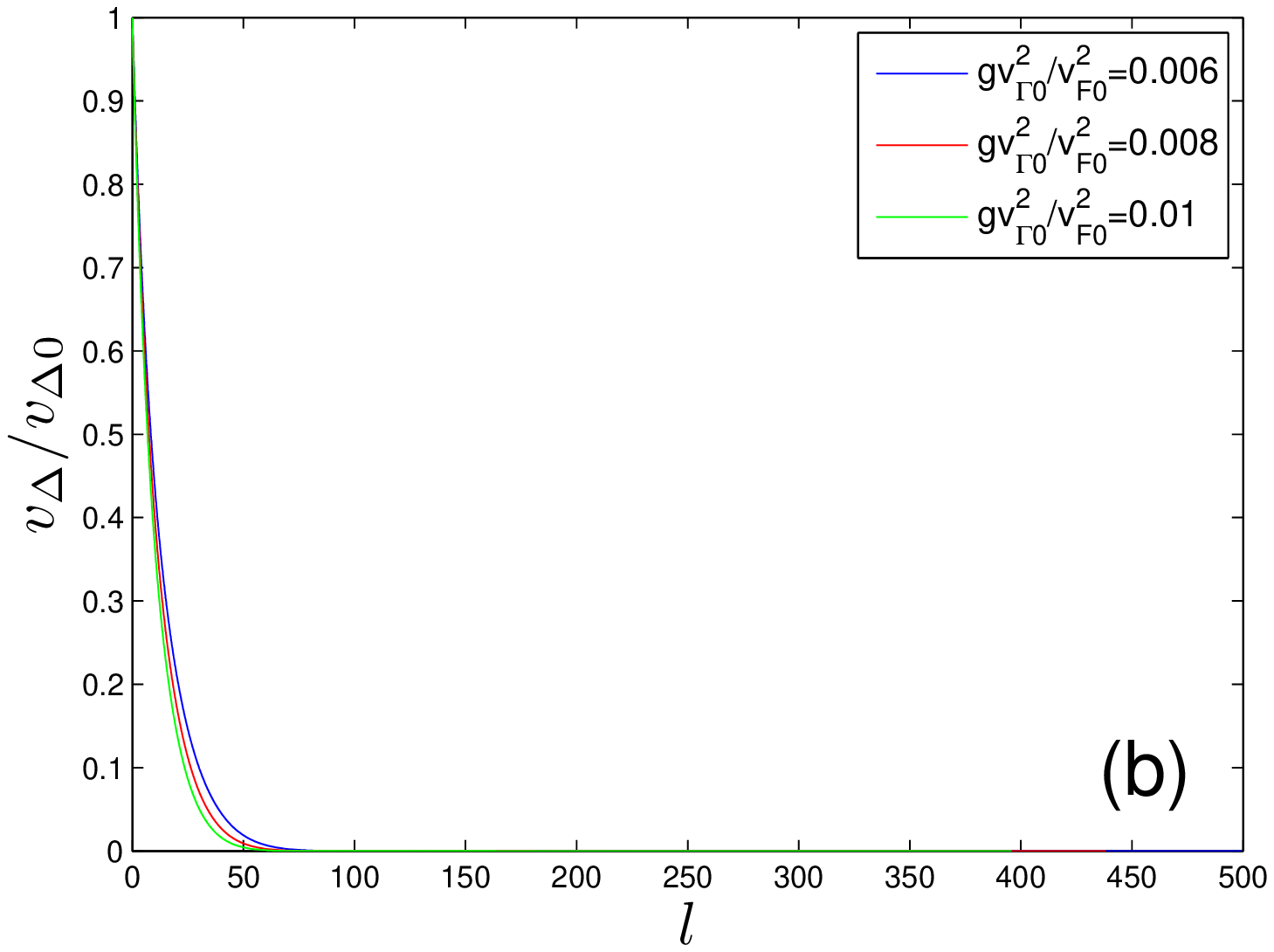}
\includegraphics[width=2.7in]{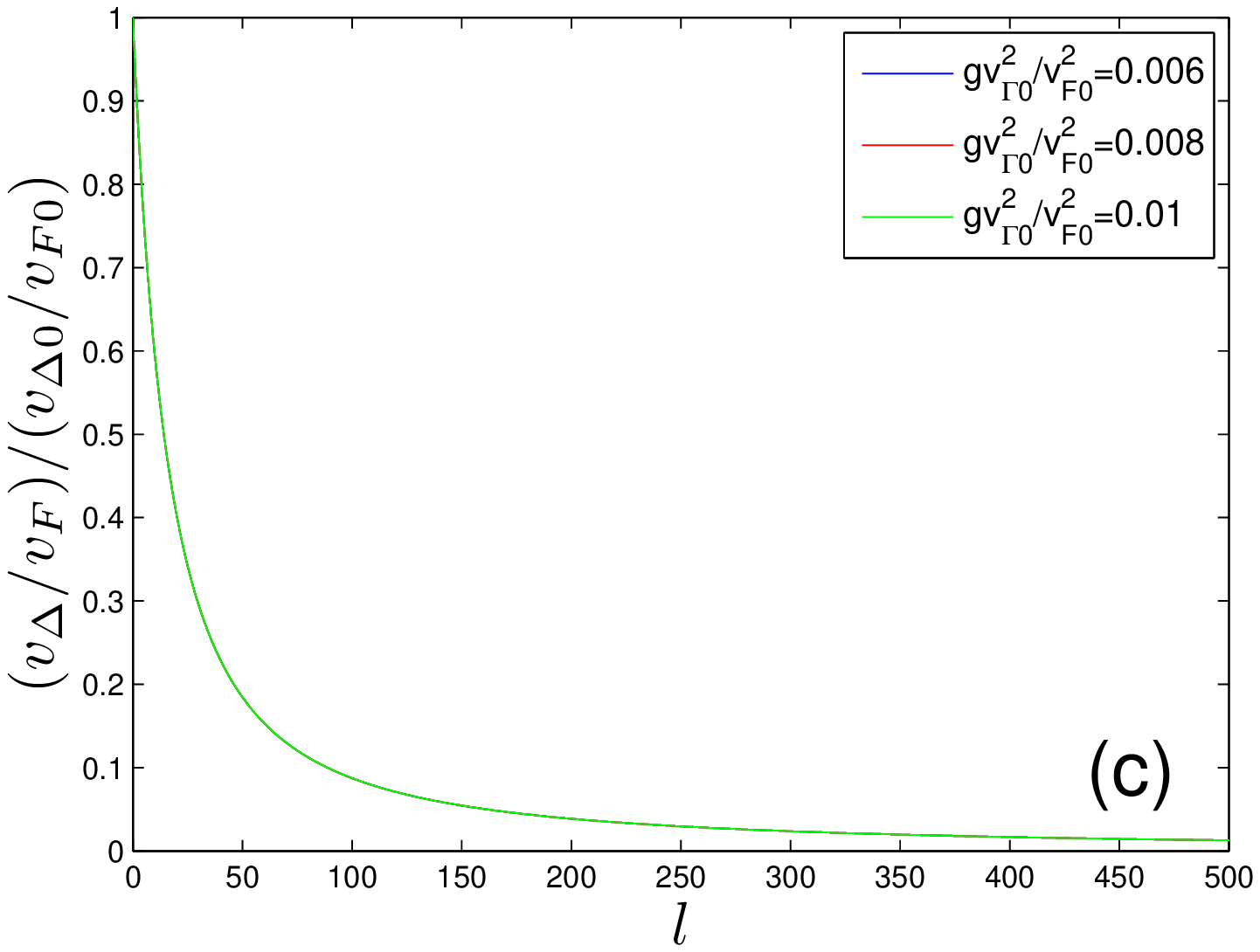}
\hspace{1.5cm}
\includegraphics[width=2.7in]{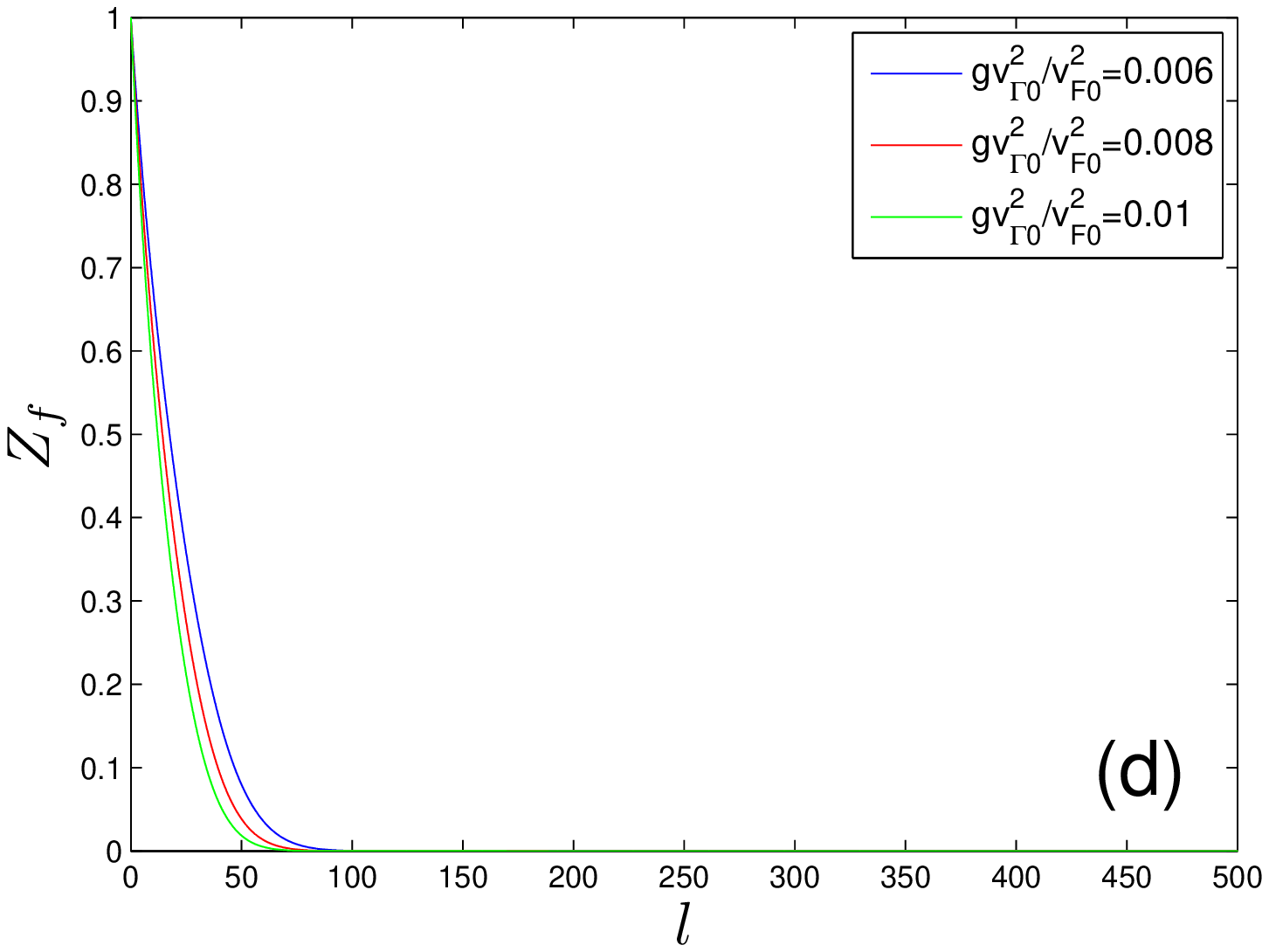}
\includegraphics[width=2.7in]{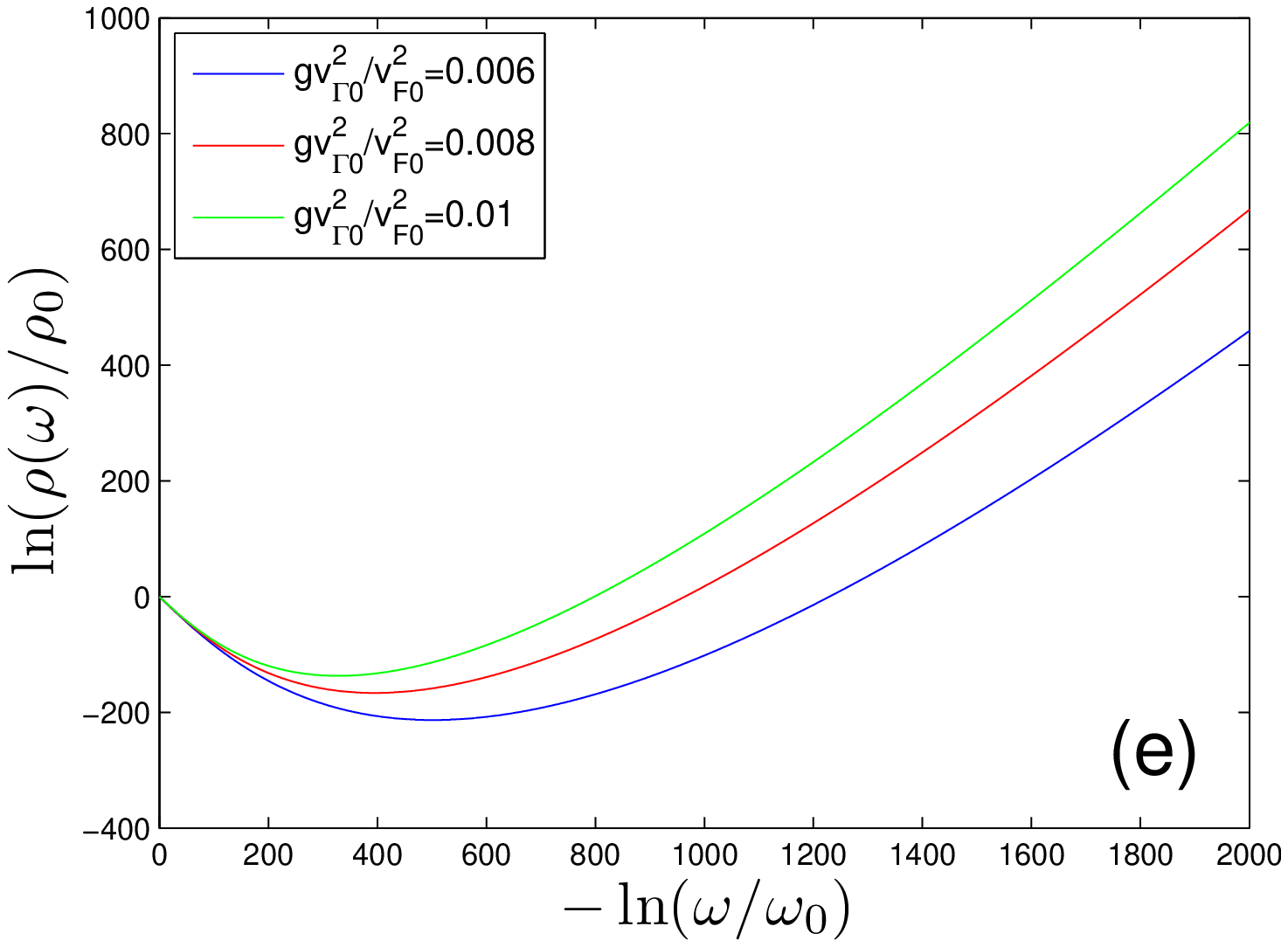}
\hspace{1.5cm}
\includegraphics[width=2.7in]{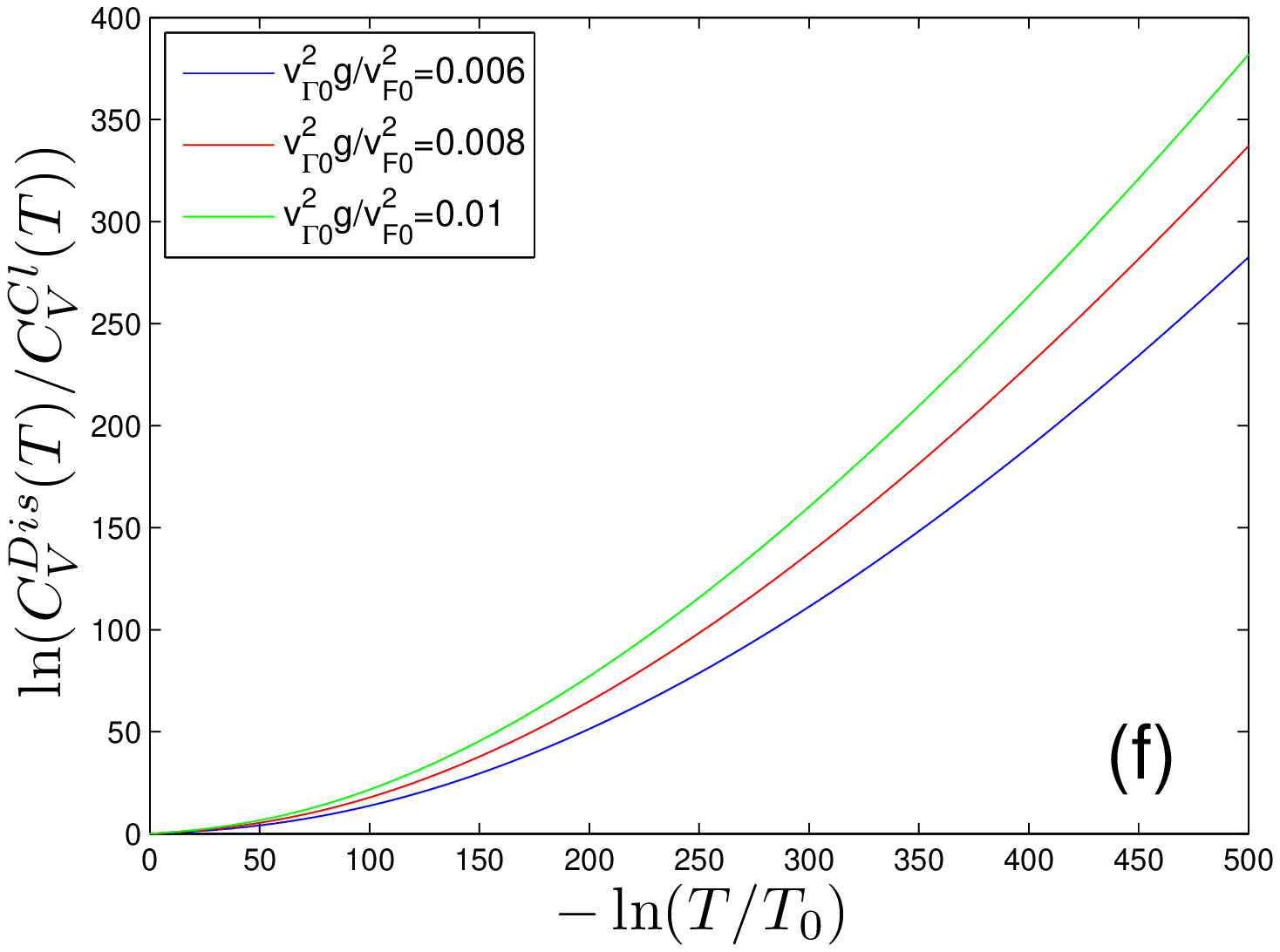}
\caption{Flows of the quantities $v_{F}$, $v_{\Delta}$,
$v_{\Delta/v_{F}}$, $Z_{f}$, $\rho$, $C_{V}$ are shown in (a), (b),
(c), (d), (e), and (f) respectively at the initial value
$v_{\Delta0}/v_{F0}=0.075$ in the presence of random gauge
potential. We have chosen $v_{\Gamma0}^{2}g/v_{F0}^{2} = 0,006,
0.008, 0.01$ and $v_{\Gamma10} = v_{\Gamma20} = v_{\Gamma0}$ for
random gauge potential. \label{Fig:VRGGP}}
\end{figure*}

\subsection{Flow of effective disorder strength}

In order to make a direct comparison to the clean case in which $C_g
= 0$ and to explicitly see the influence of a nonzero $C_g$ on the
running behaviors of various parameters, we plot the $l$-dependence
of $v_{F}$, $v_{\Delta}$, $v_{\Delta}/v_{F}$, and $Z_{f}$ in
figures~\ref{Fig:VRGGP}(a)-(d). Moreover, we show the
$\omega$-dependence of $\rho(\omega)$ in figure~\ref{Fig:VRGGP}(e) and
the $T$-dependence of $C_{V}$ in figure~\ref{Fig:VRGGP}(f),
respectively. We will discuss under what circumstances these RG
results are modified in the next subsection.

Comparing figures~\ref{Fig:RhoCl}(a) and (b) with figures
\ref{Fig:VRGGP}(a) and (b), we see that the detailed $l$-dependence
of $v_{F}$ and $v_{\Delta}$ are both altered dramatically by the
random gauge potential. As shown in figure~\ref{Fig:RhoCl}(c) and
figure~\ref{Fig:VRGGP}(c), the velocity ratio $v_{\Delta}/v_{F}$
exhibits exactly the same $l$-dependence in the clean and disordered
cases, which originates from the fact that $C_g$ does not enter into
the RG equation of velocity ratio $v_{\Delta}/v_{F}$. According to
figure~\ref{Fig:VRGZf}(a) and figure~\ref{Fig:VRGGP}(d), the
renormalization factor $Z_{f}$ flows to zero in the disordered case
much more rapidly than the clean case.

As shown in figure~\ref{Fig:VRGGP}(e), the DOS $\rho(\omega)$ is
divergent in the lowest energy limit due to random gauge potential.
This is completely different from the behaviors of clean case
presented in figure~\ref{Fig:RhoCl}. In figure~\ref{Fig:VRGGP}(f), we
plot the ratio between $C_{V}^{\mathrm{Dis}}/C_{V}^{\mathrm{Cl}}$,
where $C_{V}^{\mathrm{Dis}}$ and $C_{V}^{\mathrm{Cl}}$ are the
specific heat obtained in disordered and clean cases, respectively.
Since the solutions of the RG equations about DOS and specific heat
are modified substantially in the disordered case, it turns out that
random gauge potential is a relevant perturbation in the present
system. To verify the relevance of random gauge potential, we should
appeal to the RG analysis of the effective disorder strength.

In the action term $S_{\mathrm{dis}}$ given in equation
(\ref{Eq:ActionDisorder}), the parameters that characterize the
fermion-disorder coupling seem to be $v_{\Gamma1}$ and
$v_{\Gamma2}$. It can be seen from the RG equations that
$v_{\Gamma1}$ flows in precisely the same way as $v_{F}$, and that
$v_{\Gamma2}$ as $v_{\Delta}$. Thus, both $v_{\Gamma1}$ and
$v_{\Gamma2}$ are strongly renormalized and driven to vanish as $l
\rightarrow +\infty$. However, this does not mean that the disorders
can be simply neglected. Indeed, whether disorders are important is
determined by the ratio between the interaction energy given by
$S_{\mathrm{dis}}$ and the fermion energy $E_{k} \propto
\left(v_{F}^{2}k_{x}^{2} + v_{\Delta}^{2}k_{y}^{2}\right)^{1/2}$. We
can see that this ratio is defined by $C_{g}$, which enters into the
RG equations for the parameters $v_{F}$, $v_{\Delta}$,
$v_{\Gamma1}$, and $v_{\Gamma2}$. The effective strength of random
gauge potential should be measured by $C_{g}$, rather than
$v_{\Gamma1}$ and $v_{\Gamma2}$. Recall that $C_{g}$ is a function
of five parameters, i.e., $v_{F}$, $v_{\Delta}$, $v_{\Gamma1}$,
$v_{\Gamma2}$, and $g$. Among these parameters, $g$ is assumed to be
a dimensionless constant, but the other four parameters flow
strongly with the varying $l$. Detailed RG analysis revealed that
$C_g$ goes to infinity, namely
\begin{eqnarray}
C_g &=& \frac{g}{2\pi}\left(\frac{v_{\Gamma10}^{2}}{v_{F0}^{2}}
\frac{v_{F}}{v_{\Delta}} + \frac{v_{\Gamma20}^{2}}{v_{\Delta0}^{2}}
\frac{v_{\Delta}}{v_{F}}\right) \rightarrow +\infty,
\end{eqnarray}
in the limit $l \rightarrow +\infty$ as a consequence of the
singular renormalization of the fermion velocities ratio
$v_{\Delta}/v_{F}$. More quantitatively, the large scale behavior of
$C_{g}$ can be described by
\begin{eqnarray}
C_g \sim c_{3}l\ln l,
\end{eqnarray}
where $c_{3} = \frac{4g}{\pi^3 N}
\frac{v_{\Gamma10}^{2}}{v_{F0}^{2}}$.

It is necessary to explain here why the effective strength of random
gauge potential should be represented by $C_{g}$, rather than solely
by the coefficients $v_{\Gamma i}$ with $i = 1,2$, and why $C_{g}
\rightarrow \infty$ in the lowest energy limit. In an interacting
fermion system, the effective strength of the interaction is
characterized by the ratio between the interaction energy scale and
the kinetic energy of fermions. This ratio is widely used in
condensed matter physics to judge whether an interacting fermion
system can be defined as a strongly correlated system or not. For
instance, the normal metal with a high density of itinerant
electrons is believed to be a weakly interacting system since the
energy scale of Coulomb potential is much smaller than the Fermi
energy. In a massless Dirac fermion system (such as graphene), the
effective strength of long-range Coulomb interaction is defined as
$\alpha\sim\frac{e^2}{v_{F}}$, where $e^2$ appears in the action of
the Coulomb interaction as a coupling coefficient and the fermion
velocity $v_{F}$ reflects the order of the kinetic energy
\cite{Kotov12, Hofmann14}. Another example comes from the effective
BCS model of Dirac fermion systems, where the effective strength of
pairing interaction is characterized by $g\sim \frac{u}{v_{F}}$ with
$u$ being the coupling coefficient of pairing interaction and
$v_{F}$ fermion velocity \cite{Kopnin08, Nandkishore13}.

This criterion also applies to disordered systems. When massless
Dirac fermions couple to random gauge potential, the effective
strength of random gauge potential should be defined by $C_{g} \sim
\frac{v_{\Gamma i}^2}{v_{F}v_{\Delta}}$, rather than the coefficient
of fermion-disorder coupling $v_{\Gamma i}$ \cite{Nersesyan94,
Nersesyan95}. When $v_{\Gamma i}$ flows to zero in the lowest energy
limit, $C_{g}$ does not necessarily vanish since there is a
possibility that $v_{F}$ and $v_{\Delta}$ may vanish more (or
equally) rapidly than $v_{\Gamma i}$. When the nodal fermions couple
to both the quantum fluctuation of nematic order and random gauge
potential, the four parameters $v_{F}$, $v_{\Delta}$, $v_{\Gamma1}$,
and $v_{\Gamma2}$ all flow to zero in the lowest energy limit, but
the effective strength of random gauge potential becomes very large,
namely $C_{g} \rightarrow \infty$. This originates from the fact
that $\frac{v_{\Gamma1}}{v_{F}} = \frac{v_{\Gamma10}}{v_{\Gamma20}}$
and $\frac{v_{\Gamma2}}{v_{\Delta}} =
\frac{v_{\Gamma20}}{v_{\Delta0}}$ are constants, while at the same
time the velocity ratio $v_{\Delta}/v_{F} \rightarrow 0$. The
behavior $C_{g}\rightarrow \infty$ at low energies indicates that
random gauge potential is relevant. To see this point, we neglect
the nematic order and consider only the coupling of nodal fermions
to random gauge potential, which results in simplified RG equations:
\begin{eqnarray}
\fl \frac{dv_{F}}{dl} = -C_{g}v_{F},\,\frac{dv_{\Delta}}{dl} =
-C_{g}v_{F},\,\frac{d(v_{F}/\Delta)}{dl} =
0,\,\frac{dv_{\Gamma1}}{dl} = -C_{g}v_{F},\, \frac{dv_{\Gamma2}}{dl}
= -C_{g}v_{F}.
\end{eqnarray}
Using the two relations $\frac{v_{\Gamma1}}{v_{F}} =
\frac{v_{\Gamma10}}{v_{\Gamma20}}$ and
$\frac{v_{\Gamma2}}{v_{\Delta}} = \frac{v_{\Gamma20}}{v_{\Delta0}}$,
it is easy to show that
\begin{eqnarray}
C_{g} = \frac{g}{2\pi}\frac{v_{\Gamma1}^{2} +
v_{\Gamma2}^{2}}{v_{F}v_{\Delta}} =
\frac{g}{2\pi}\frac{v_{\Gamma10}^{2} +
v_{\Gamma20}^{2}}{v_{F0}v_{\Delta0}} = C_{g0}.
\end{eqnarray}
In this case, $v_{F}$, $v_{\Delta}$, $v_{\Gamma1}$, and
$v_{\Gamma2}$ depend on length scale $l$ as
\begin{eqnarray}
\frac{v_{F}}{v_{F0}} = \frac{v_{\Delta}}{v_{\Delta0}} =
\frac{v_{\Gamma1}}{v_{\Gamma10}} = \frac{v_{\Gamma2}}{v_{\Gamma20}}
= e^{-C_{g0}l},
\end{eqnarray}
which all vanish rapidly in the limit $l \rightarrow \infty$. Since
$C_{g}$ is a constant, random gauge potential is marginal. From the
above calculations, we can conclude that the behavior $C_{g}
\rightarrow \infty$ obtained in the presence of both nematic
fluctuation and random gauge potential is directly related to the
extreme velocity anisotropy $v_{\Delta}/v_{F} \rightarrow 0$ induced
by the quantum fluctuation of nematic order. The flow of $C_{g}$
towards strong coupling regime is a clear signature that random
gauge potential should have substantial physical effects on the
low-energy behaviors of nodal fermions, which will be discussed in
the next subsection.

We notice other interesting correlated electron models in which the
coupling coefficient of an interaction vanishes at low energies, but
the interaction is not negligible due to the more (or equally) rapid
decrease of the kinetic energy of electrons. Recently, Sur and Lee
studied the influence of quantum fluctuations of an
antiferromagnetic (AF) order at an AF quantum critical point in a
metal supporting one-dimensional Fermi surface \cite{Sur15}. In
particular, they showed that the coupling coefficient of the
interaction flows to zero at low energies. However, the fermion
velocity also vanishes, thus the interaction cannot be simply
neglected. Actually, Sur and Lee found that the interaction drives
the system to become a so-called quasi-local strange metal that is
apparently qualitatively different from the free fermion system.

\subsection{Physical effects of random gauge potential}

How should we understand the divergence of $C_{g}$? In order to
answer this question, we first consider the non-interacting system
that contains only nodal fermions and random gauge potential. In
this case, the RG equation of fermion DOS becomes
\begin{eqnarray}
\frac{d\ln\rho}{d\ln\omega} = \frac{1  - C_{g0}}{1 +
C_{g0}},\label{Eq:RhoRGEqNoInteraction}
\end{eqnarray}
with
\begin{eqnarray}
C_{g0} = \frac{g}{2\pi}\frac{v_{\Gamma10}^{2} +
v_{\Gamma20}^{2}}{v_{F0}v_{\Delta0}}.\label{Eq:CgNoInteraction.}
\end{eqnarray}
If $C_{g0} < 1$, we have
\begin{eqnarray}
\rho(\omega)\sim \omega^{\alpha}\quad \mathrm{with}\quad \alpha =
\frac{1 - C_{g0}}{1 + C_{g0}},\label{Eq:DOSOnlyDisorder}
\end{eqnarray}
where $\alpha$ satisfies $0 < \alpha < 1$. In this case, the RG
equation for specific heat is given by
\begin{eqnarray}
\frac{d\ln C_{V}}{d\ln T}=2-\frac{2C_{g0}}{1+C_{g0}}.
\end{eqnarray}
The solution of this equation is given by
\begin{eqnarray}
C_{V}(T)\sim T^{\beta}\quad \mathrm{with}\quad \beta = 2 -
\frac{2C_{g0}}{1 + C_{g0}}.
\label{Eq:SpecificHeatOnlyDisorder}
\end{eqnarray}
If $C_{g} > 1$, it is easy to get
\begin{eqnarray}
\rho(\omega) \rightarrow \infty
\end{eqnarray}
in the limit $\omega \rightarrow 0$. The divergence $\rho(\omega)$
indicates the emergence of a disorder dominated diffusive state, in
which a finite disorder scattering rate $\gamma$ and a finite
zero-energy DOS $\rho(0)$ are generated \cite{Altland02,
Chakravarty, Roy}. The value $C_{g} = 1$ defines a QCP for the
quantum phase transition between the ballistic and diffusive states
of nodal fermions. Therefore, a weak random gauge potential gives
rise to power law behavior of $\rho(\omega)$, whereas a sufficiently
strong disorder can trigger the quantum phase transition to a
diffusive state.

In the presence of both nematic critical fluctuation and random
gauge potential, the fact that $C_{g}\rightarrow \infty$ in the
lowest energy limit signals the development of a disorder dominated
diffusive state and the generation of a finite $\gamma$ and a finite
$\rho(0)$ even in the case of weak random gauge potential. Although
the perturbative RG method provides a powerful tool to judge whether
and when a phase transition takes place, it cannot be used to
compute the precise value of $\gamma$. To calculate $\gamma$, one
usually needs to construct a self-consistent equation for the
retarded fermion self-energy by properly considering both the
fermion-disorder scattering and the influence of quantum critical
fluctuation of nematic order \cite{Balatsky06}. This is an
interesting yet complicated issue, which is beyond the scope of the
present work and subjected to future research. Here, we use the RG
method to make a rough estimation for the energy scale of $\gamma$.
As shown in Fig.~\ref{Fig:VRGGP}(e), for small given values of
$C_{g0}$ and $v_{\Delta0}/v_{F0}$, the solution of the RG equation
of DOS should have the following properties: $\rho(\omega)$
decreases as the varying energy scale decreases, but tends to
increase when the energy scale exceeds a critical value
$E_{c}(C_{g0},v_{\Delta0}/v_{F0})$, which is a function of $C_{g0}$
and $v_{\Delta0}/v_{F0}$. The magnitude of $\gamma$ should be an
increasing function of $E_{c}$. In the following calculations of
$\rho(0)$ and $C_V(T)$, we will regard $\gamma$ as an undetermined
constant. Fortunately, the qualitative behaviors of $\rho(0)$ and
$C_{V}(T)$ in the low energy regime is independent of the precise
value of $\gamma$.

The imaginary part of retarded fermion self-energy can be
generically written as
\begin{eqnarray}
\mathrm{Im}\Sigma^R(\omega) \approx \gamma +
\mathrm{Im}\Sigma_{\mathrm{nem}}^R(\omega), \label{Eq:DampingAll}
\end{eqnarray}
where $\mathrm{Im}\Sigma_{\mathrm{nem}}^R(\omega)$ is the
contribution induced solely by the nematic order. The
disorder-induced scattering rate $\gamma$ represents a
characteristic energy scale. At energies higher than $\gamma$,
namely $\omega > \gamma$,
$\mathrm{Im}\Sigma_{\mathrm{nem}}^R(\omega)$ dominates over $\gamma$
and all the RG results for $v_{F}$, $v_{\Delta}$, $Z_{f}$,
$\rho(\omega)$, and $C_{V}(T)$ shown in figures~\ref{Fig:VRGGP}(a)-(f)
are still applicable. At $\omega <\gamma$, the diffusive motion of
nodal fermions and its interplay with critical nematic fluctuation
govern the low-energy properties of the system.

Once a finite $\gamma$  is generated, the renormalized velocities
$v_{F}$ and $v_{\Delta}$ no more vanish at low energies, which is
apparently different from the clean case. Instead, as the energy
scale decreases, both $v_{F}$ and $v_{\Delta}$ are saturated to
certain finite values, denoted by $v_{F0}'$ and $v_{\Delta0}'$,
below the energy scale set by $\gamma$. Hence, there is no extreme
velocity anisotropy in the diffusive state. The fermion DOS and
specific heat also exhibit different behaviors comparing to the
clean case. To demonstrate the difference in DOS, we write the
retarded propagators of nodal fermions in the forms:
\begin{eqnarray}
G_{1a}^{R}(\omega,\mathbf{k}) &=& \frac{1}{-(\omega+i\gamma) +
v_{F0}'k_{x}\tau^{z} + v_{\Delta0}'k_{y}\tau^{x}},
\\
G_{2a}^{R}(\omega,\mathbf{k}) &=& \frac{1}{-(\omega+i\gamma) +
v_{F0}'k_{y}\tau^{z} + v_{\Delta0}'k_{x}\tau^{x}}.
\end{eqnarray}
Calculations find that the fermion DOS depends on $\omega$ as
\begin{eqnarray}
\fl \rho(\omega) &=& \frac{N\gamma}{2\pi v_{F0}'v_{\Delta0}'}
\left[\ln \left(\frac{\left(\gamma^2 - \omega^2 +
v_{F0}'v_{\Delta0}'\Lambda^{2}\right)^2 +
4\omega^2\gamma^2}{\left(\gamma^2-\omega^2\right)^2 +
4\omega^2\gamma^2}\right)\right.\nonumber \\
\fl &&\left.+\frac{2\omega}{\gamma} \arctan\left(\frac{\gamma^2 -
\omega^2 + v_{F0}'v_{\Delta0}'\Lambda^2}{2\omega\gamma}\right) -
\frac{2\omega}{\gamma}\arctan\left(\frac{\gamma^2 - \omega^2}
{2\omega\gamma}\right)\right].
\end{eqnarray}
In the case $\sqrt{v_{F0}'v_{\Delta0}'}\Lambda \gg \gamma$, we have
\begin{eqnarray}
\rho(0)\approx&\frac{2N\gamma}{\pi v_{F0}'v_{\Delta0}'}
\ln\left(\frac{\sqrt{v_{F0}'v_{\Delta0}'}\Lambda}{\gamma}\right).
\end{eqnarray}
In order to compute the specific heat, it is convenient to invoke
the standard Matsubara formalism of fermion propagators, i.e.,
\begin{eqnarray}
G_{1a}(\omega_{n},\mathbf{k}) &=& \frac{1}{-i(\omega_{n} +
\gamma\mathrm{sgn}(\omega_{n})) + v_{F0}'k_{x}\tau^{z} +
v_{\Delta0}'k_{y}\tau^{x}},
\nonumber \\
G_{2a}(\omega_{n},\mathbf{k}) &=& \frac{1}{-i(\omega_{n} +
\gamma\mathrm{sgn}(\omega_{n})) + v_{F0}'k_{y}\tau^{z} +
v_{\Delta0}'k_{x}\tau^{x}}, \nonumber
\end{eqnarray}
where $\omega_{n}=(2n+1)\pi T$ with $n$ being integer.
The fermion free energy is given by
\begin{eqnarray}
F = -\frac{2NT}{v_{F0}'v_{\Delta0}'} \sum_{\omega_{n}}\int
\frac{d^2k}{(2\pi)^2}\ln\left[\left(\omega_{n} +
\gamma\mathrm{sgn}(\omega_{n})\right)^2+k^2\right].\nonumber
\end{eqnarray}
Summing over $\omega_n$ leads to
\begin{eqnarray}
\fl F &\approx& -\frac{4N}{\pi v_{F0}'v_{\Delta0}'}
\int\frac{d^2k}{(2\pi)^2}\left\{k\arctan\left(\frac{k}{\pi T +
\gamma}\right)+\frac{\gamma}{2}\ln\left[\left(\pi T + \gamma\right)^{2} +
k^{2}\right]\right\}.
\end{eqnarray}
Now the fermion specific heat in the low energy regime can be approximately given by
\begin{eqnarray}
C_{V} &\approx& \frac{6NT}{v_{F0}'v_{\Delta0}'}\gamma
\ln\left(\frac{\Lambda}{\gamma}\right),\label{Eq:CvDisorder}
\end{eqnarray}
which depends on $T$ linearly. We can see that DOS and specific heat
obtained in the diffusive state exhibit entirely different behaviors
from the unconventional non-Fermi liquid state below the energy
scale $\gamma$.

We finally make a brief remark on the behavior of the system staying
away from the nematic QCP. Suppose the system stays at a distance
$r$ to the QCP, the RG results are still valid and the fermion
velocity ratio is still renormalized at energies higher than the
scale corresponding to $r$. However, the renormalization terminates
at certain low energy scale. Therefore, the effective strength of
random gauge potential is moderately enhanced compared its bare
value, though $C_g$ does not diverge.

\section{Comparison with experiments\label{Sec:CompaisionExperiments}}

In this section, we address the possible connection between the
theoretical results obtained in the last sections and the
phenomenology of cuprate superconductors. We are particularly
interested in three existing experimental findings about some of the
unusual properties of the superconducting dome.

\subsection{Anomalous residual linear-$T$ term of specific heat in cuprates}

We first discuss the residual specific heat in cuprates. Due to the
line nodes of $d$-wave gap, the specific heat in the superconducting
phase of cuprates is expected to exhibit a $T^2$ behavior, i.e.,
$C_V(T)\propto T^2$, at low $T$. This expectation is generically
consistent with experiments \cite{Fisher07}. In the lowest $T$
limit, experiments have observed a residual linear $T$ term of $C_V(T)$
\cite{Fisher07, Revaz98, Wright99, Riggs11}, which is usually
attributed to the finite fermion DOS generated by disorder
scattering and is also well consistent with the result given by
equation (\ref{Eq:CvDisorder}). There is, however, an unexpected
experimental finding \cite{Fisher07, Riggs11} that the residual
specific heat of YBCO is obviously larger in magnitude than that of
La$_{2-x}$Sr$_x$CuO$_4$ (LSCO), although the former material is
known to be cleaner than the latter. Apparently, disorder scattering
alone is not capable of accounting for this experimental fact.
Recently, coexistence of $d$-wave superconductivity with a loop
current order was proposed to give a possible explanation
\cite{Allais12, Kivelson12, WangVafek13} for the large residual
linear-$T$ term of specific heat in YBCO. In this scenario, when
$d$-wave superconductivity coexists with a loop current order, two
of four nodal points are converted to finite Fermi pockets of
Bougoliubov quasiparticles, which then generates a finite $\rho(0)$
and a residual linear-$T$ term of specific heat \cite{Allais12,
Kivelson12, WangVafek13, Berg08}. The recent ultrasound measurements
showed the possible evidence for the existence of loop current
order in YBCO\cite{Shekhter13,Zaanen13}.

Here we propose an alternative explanation for the above anomalous
experimental results of residual specific heat. Our RG analysis
found that the effective strength of random gauge potential, being
the most relevant disorder to cuprates \cite{Nersesyan95}, is
strongly enhanced by the critical nematic fluctuation, which in turn
increases the residual value of the specific heat. To understand the
role of quantum nematic fluctuation, we firstly consider only the
coupling of nodal fermions to random gauge potential. In this
special case, $C_{g}$ does not flow and thus remains a constant,
namely $C_{g} = C_{g0}$. If $C_{g0}$ is very small, which means the
system is only slightly disordered, the behavior of $\rho(\omega)$
and specific heat would be governed by equations
(\ref{Eq:DOSOnlyDisorder}) and (\ref{Eq:SpecificHeatOnlyDisorder}),
respectively. In this case, the system does not develop a finite
$\rho(0)$ and there is no residual liner-$T$ term of specific heat.
If the system is quite disordered such that $C_{g0} > 1$, it enters
a diffusive state with a finite scattering rate $\gamma$, which
induces a finite $\rho(0)$ and a residual linear-$T$ term of
specific heat. Apparently, the residual specific heat is larger in
more disordered systems, which is not consistent with the
aforementioned experiments of residual specific heat. If we consider
both the quantum fluctuation of nematic order and random gauge
potential, $C_{g}$ is significantly enhanced and flows to strong
coupling at low energies even if $C_{g0}$ takes an arbitrarily small
value. This implies that a cleaner compound might acquire a larger
amount of $\gamma$ and naturally a larger $\rho(0)$.

To make a more careful comparison between theories and experiments,
we now briefly discuss the doping dependence of our results. We use
$\delta$ to denote the doping concentration and $\delta_c$ the
nematic QCP. At zero temperature, the mass of nematic field $\phi$
is proportional to the difference between $\delta$ and $\delta_c$,
namely $r \sim |\delta - \delta_{c}|$. When the cuprate is at a
distance $r$ away from the nematic QCP, the quantum fluctuation of
nematic order is not critical, but remains important for small $r$.
At the energy scales larger than $r$, the fermion velocity
anisotropy is still considerably enhanced, which leads $C_{g}$ to
flow to larger values. For a given small value of $C_{g0}$, $C_{g}$
can flow to a sufficiently large value to induce a diffusive state
and generate a finite scattering rate $\gamma$, provided that $r$ is
made sufficiently small. In this case, the quantum fluctuation of
nematic order can result in a finite $\rho(0)$ and a finite residual
linear-$T$ term of specific heat. If the bare value $C_{g0}$ is
large enough, random gauge potential itself suffices to generate a
finite $\gamma$. Including the quantum fluctuation of nematic order
leads to larger values of both $C_{g}$ and $\gamma$. In any case,
the quantum fluctuation of nematic order tends to amplify $\gamma$,
which naturally increase $\rho(0)$ and the residual specific heat.

A number of recent experiments provided strong evidence supporting
the existence of nematic order in YBCO, whereas there is little
evidence for nematicity in LSCO. Although YBCO is cleaner than LSCO,
the quantum fluctuation of nematic order in the former
superconductor can drive $C_{g}$ to flow to much larger values. As a
consequence, the residual linear-in-$T$ term of specific heat of
YBCO would be larger than that of LSCO. We emphasize that, disorder
itself cannot explain the anomalous behaviors of the residual
specific heat observed in YBCO and LSCO, and it is necessary to
consider the interplay of quantum nematic fluctuation and random
gauge potential.

For the above elaboration, we know that the roles played by the
quantum nematic fluctuation and random gauge potential depends on
doping $\delta$. To simplify discussion, we assume the bare value
$C_{g0}$ displays only a weak $\delta$-dependence. Since the quantum
fluctuation of nematic order is most pronounced at the nematic QCP,
the scattering rate $\gamma$ and consequently the coefficient of the
residual liner-$T$ term of specific heat are maximal at the QCP, and
decrease as the system moves away from this QCP. This doping
dependence is observable, and can be examined by experiments. Within
the loop current order scenario, both the zero-energy DOS $\rho(0)$
and the residual linear-$T$ term of specific heat are proportional
to the order parameter of the loop current order, which decreases
with the growing doping in the underdoped region.

Early experiments \cite{Fisher07, Revaz98, Wright99} found that the
coefficient of linear-$T$ term of specific heat at optimally doped
YBCO is roughly $2$ mJ$\cdot$mol$^{-1}$$\cdot$K$^{-2}$. More recent
measurements performed in underdoped YBCO \cite{Riggs11} revealed
that this coefficient is $1.85\pm 0.06$
mJ$\cdot$mol$^{-1}$$\cdot$K$^{-2}$. We feel that the currently
available experimental data about the doping dependence of this
coefficient are still quite limited. We expect more extensive
measurements would be performed in the future to extract a more
quantitative doping dependence of the coefficient, which could help
to judge whether the scenario proposed in this paper works.

\subsection{Strong damping rate of nodal fermions in optimally doped BSCCO}

We next apply the RG results to understand the observed damping rate
of nodal fermions in cuprates. Valla \emph{et al.} \cite{Valla99}
have performed extensive angle resolved photoemission spectroscopy
measurements in optimally doped
Bi$_{2}$Sr$_{2}$CaCu$_{2}$O$_{8+\delta}$ (BSCCO). Their main
discovery is that the nodal fermions exhibit a MFL-type damping rate
in the normal state above $T_c$, which is in general consistency
with the observed linear resistivity. They further found
\cite{Valla99} that the linear damping rate is not sensitive to the
onset of superconductivity and persists well below $T_c$. This was
out of expectation since previous BCS weak coupling analysis
\cite{Orenstein00,Titov95} had predicted a quite weak damping rate,
i.e., $\mathrm{Im}\Sigma^R(\omega,T) \propto
\max(\omega^{3},T^{3})$, in the superconducting phase. Several
scenarios \cite{Orenstein00,Vojta00A, Khveshchenko01} were proposed
to account for the nearly MFL behavior. In particular, Vojta
\emph{et al.} \cite{Vojta00A} and Khveshchenko and Paaske
\cite{Khveshchenko01} have argued that the strong fermion damping
may arise from a secondary phase transition from a
$d_{x^2-y^2}$-wave superconducting state to a $d_{x^2-y^2}+is$ or
$d_{x^2-y^2}+id_{xy}$-wave superconducting state.

Experimentally, the existence of a nematic phase was observed in
BSCCO by Lawler \emph{et al.} \cite{Lawler10}. More recent
experimental studies of Fujita \emph{et al.} provided further
evidence pointing towards the existence of a nematic QCP in the
vicinity of optimal doping in BSCCO \cite{Fujita14, Fradkin15}.
Therefore, it seems natural to account for the experimental finding
of Valla \emph{et al.} by considering the quantum critical
fluctuation of nematic order. We have showed through RG analysis
that the nodal fermion damping rate caused by critical nematic
fluctuation, described by $\mathrm{Im}
\Sigma_{\mathrm{nem}}^{R}(\omega)$, is slightly weaker than that of
a MFL. In realistic experiments, it is difficult to distinguish this
non-Fermi liquid state from a MFL state. Therefore, presence of a
nematic QCP provides an alternative scenario for the nearly MFL
behavior observed in optimally doped BSCCO. However, the nearly MFL
behavior occurs only at energies higher than the value
$\omega_{\gamma}$ set by disorder scattering rate $\gamma$. Indeed,
at $\omega < \omega_{\gamma}$, $\gamma$ dominates over
$\mathrm{Im}\Sigma_{\mathrm{nem}}^{R}(\omega)$, thus a finite
zero-energy DOS $\rho(0)$ is generated. To summarize, our RG results
are qualitatively consistent with the quasiparticle self-energy
$\mathrm{Im}\Sigma(\omega,T) = \gamma+\max(\omega,T)$ observed in
reference \cite{Valla99}.

\subsection{Temperature dependence of fermion velocity in BSCCO}

Another experiment is the observation of Plumb \emph{et al.}
\cite{Plumb10} that the fermion velocity $v_{F}$ along the nodal
directions increases as $T$ grows in BSCCO. We know from RG results
that the critical nematic fluctuations drive the fermion velocities
to vanish in the lowest energy limit. This means the velocities must
increase as the energy scale is growing. Therefore, the nematic
quantum fluctuation will induce increment of fermion velocities when
the thermal energy increases with growing $T$, which is
qualitatively well consistent with the observation of Plumb \emph{et
al}. Technically, one can translate the $l$-dependence of fermion
velocity to a $T$-dependence by making the transformation $T = T_0
e^{-l}$ \cite{She15, WangLiu15}. With the help of this
transformation, it is easy to show that the fermion velocity $v_F$
is an increasing function of $T$. Therefore, the singular
renormalization of fermion velocities of nodal fermions induced by
the quantum fluctuation of namatic order, which is first discovered
by Huh and Sachdev \cite{Huh08}, agrees with the $T$-dependence of
$v_{F}$ observed in reference \cite{Plumb10}.

\section{Summary\label{Sec:Summary} and discussions}

In summary, we have found that the nodal fermions of $d$-wave
superconductors constitute an unconventional non-Fermi liquid, which
exhibits a weaker violation of Fermi liquid description than a MFL,
due to the quantum critical fluctuation of nematic order. This
unusual state represents a novel quantum state of matter that cannot
be well captured by the traditional classification of (non-) Fermi
liquids and thus enriches our knowledge of strong electron
correlation effects. We also have calculated the fermion DOS and
specific heat after incorporating the unusual renormalization of
fermion velocities. When a gauge-potential-type disorder is added to
the system, we have analyzed its interplay with the quantum
fluctuation of nematic order, and found that the effective disorder
strength flows to strong coupling, leading to diffusive motion of
nodal fermions. Therefore, even a weak random gauge potential can
drive a quantum phase transition from a unconventional non-Fermi
liquid state to a disorder dominated diffusive state. However, the
unusual fermion damping induced by the nematic order is more
important than the disorder scattering at high temperatures, where
the nodal fermions still display the unconventional non-Fermi liquid
behaviors. We finally have discussed the connection between our
theoretic results and a number of interesting experiments in the
context of some cuprate superconductors.

We now would like compare our work to the existing extensive works
on the non-Fermi liquid behaviors in two-dimensional metals produced
by the quantum critical fluctuation of nematic order. At the QCP of
Pomeranchuk instability in two-dimensional metals, the quantum
fluctuation of nematic order can lead to very strong fermion damping
\cite{Oganesyan01, Metzner03, DellAnna06, Rech06,
Garst10,Metlitski10}. To the leading order, it is found
\cite{Oganesyan01, Metzner03, DellAnna06, Rech06, Garst10,
Metlitski10} the fermion damping rate behaves as
$\mathrm{Im}\Sigma(\omega) \sim \omega^{2/3}$ and the quasiparticle
residue $Z_{f}\sim\omega^{1/3}$. Since $Z_f$ vanishes in the limit
$\omega \rightarrow 0$, this QCP exhibits non-Fermi liquid behavior.
In this paper, we have considered the interaction between the
quantum fluctuation of nematic order and massless nodal fermions in
the superconducting dome of cuprate superconductors. It is apparent
that the quantum critical fluctuation of nematic order gives rise to
a stronger fermion damping effect in metals than in the
superconducting dome of cuprates. This difference should be owing
to the different forms of the kinetic energies of fermionic
excitations. In the context of cuprates, the kinetic energy of the
massless Dirac fermions excited from the superconducting gap nodes
is $E = \sqrt{v_{F}^{2}k_{x}^{2} + v_{\Delta}^{2}k_{y}^{2}}$. In
contrast, in a two-dimensional metal with a finite Fermi surface,
the kinetic energy of fermions can be written as $E = v_{F}k_{x} +
\frac{k_{y}^{2}}{2m}$, where $k_{x}$ is the momentum component
perpendicular to the Fermi surface and $k_{y}$ is the momentum
component along the tangential direction \cite{Oganesyan01,
Metzner03, DellAnna06, Rech06, Garst10, Metlitski10}. In the
low-energy regime, the latter kinetic energy is smaller that the
former for the same given values of momenta, which indicates that
the interaction plays a more important role in the latter system
than the former. To further demonstrate this difference, we consider
the different roles of the long-range Coulomb interaction in a
two-dimensional Dirac semimetal and a two-dimensional semi-Dirac
semimetal. In a Dirac semimetal, the kinetic energy of Dirac
fermions is simply $E = v_{F}k$ with $k = \sqrt{k_{x}^{2} +
k_{y}^{2}}$. RG analysis showed that the residue $Z_{f}$ approaches
a finite value at low energies, so the system is a normal Fermi
liquid\cite{Gonzalez96, Gonzalez99, Kotov12, WangLiu14, Hofmann14}.
In a semi-Dirac semimetal, the kinetic energy of fermions is written
as $E = \sqrt{v_{F}^{2}k_{x}^{2} + \frac{k_{y}^{2}}{4m^2}}$
\cite{Isobe16, Cho16}. In this case, the long-range Coulomb
interaction drives $Z_f$ to vanish in the lowest energy limit, i.e.,
$Z_{f} \rightarrow 0$, which apparently implies the breakdown of
Fermi liquid behavior \cite{Isobe16}. Once again, we see that the
ratio between the interaction energy scale and the kinetic energy is
a crucial quantity to determine the low-energy behaviors of an
interacting fermion system.

The coupling of nodal fermions to the quantum fluctuation of
antiferromagnetic (AF) order is also an interesting problem
\cite{Uemura04, Pelissetto08}. Uemura \cite{Uemura04} considered the
coupling of nodal fermions to the $(\pi,\pi)$ AF fluctuation, and
suggested a possibility that the $(\pi,\pi)$ AF fluctuation can
connect two nodal charges in different hole pockets and then
generate a bound state of two nodal charges. More recently,
Pelissetto \emph{et al.} studied a number of different couplings
between nodal fermions and AF fluctuations using RG method
\cite{Pelissetto08}. An interesting results is that, though most of
these couplings are irrelevant, there emerges a nearly marginal
coupling between nodal fermions and an effective, AF-order induced
nematic fluctuation. This nearly marginal coupling is found
\cite{Pelissetto08} to results in a fermion damping rate that is
nearly linear in energy or temperature.

The electronic nematic state has been observed not only in some
cuprate superconductors, but also in a number of iron-based
superconductors \cite{Chubukov15A}: 122 family, such as hole doped
Ba$_{1-x}$K$_{x}$Fe$_{2}$As$_{2}$, electron doped
Ba(Fe$_{1-x}$Co$_{2}$)$_{2}$As$_{2}$, and isovalent-doped
BaFe$_{2}$(As$_{1-x}$P$_{x}$)$_{2}$; 111 family, such as NaFeAs;
1111 family, such as LaFeAsO; 11 family, such as FeSe
\cite{Chubukov15A, Fernandes14}. In most of these compounds, the
nematic order emerges in accompany with a spin density wave (SDW)
order. However, there are also exceptions. For instance, the nematic
order is observed in FeSe without any evidence for SDW order
\cite{Chubukov15A, Fernandes14, Bohmer15, Baek15, Chubukov15B, Yu15,
Glasbrenner15, WangFa15, Jiang16}. Whether the nematic order
observed in iron-based superconductors is generated by the
fluctuation of SDW order or the orbital degrees of freedom is still
in fierce debate \cite{Chubukov15A, Fernandes14, Bohmer15, Baek15,
Chubukov15B, Yu15, Glasbrenner15, WangFa15, Jiang16, Lee09}. Recent
experimental studies have unambiguously showed that there is a QCP
in the superconducting dome at the optimal doping of
BaFe$_{2}$(As$_{1-x}$P$_{x}$)$_{2}$. This QCP may correspond to the
critical point for a SDW order or nematic QCP \cite{Shibauchi14,
Dioguardi16}, and is expected to exhibit rich quantum critical
phenomena. Moreover, there are clear evidences that the
superconducting gap of BaFe$_{2}$(As$_{1-x}$P$_{x}$)$_{2}$ has nodal
line points \cite{Shibauchi14}. Since the quantum fluctuation of
nematic order is peaked at zero momentum \cite{She15, Fernandes14,
Fernandes12}, the nodal fermions excited from the nodal line points
might couple strongly to the quantum fluctuation of nematic order at
the nematic QCP \cite{She15}. This coupling is physically analogous
to the model considered in this work, and it would be interesting to
study this coupling by means of RG method. The RG analysis performed
in this work could be generalized to study the possible non-Fermi
liquid behavior and disorder effects in the context of
BaFe$_{2}$(As$_{1-x}$P$_{x}$)$_{2}$ and other iron based
superconductors, where the multi-band effects and different gap
symmetry need to be seriously taken into account.

\ack{G.Z.L. and J.R.W. would like to thank Dr. Jing Wang for very
helpful discussions. We acknowledge the support by the National
Natural Science Foundation of China under Grants No.11274286,
No.11574285, No.11504379, and No.U1532267.}

\appendix

\section{Expressions of $C_{1,2,3}$\label{App:C1C2C3}}

The expressions of $C_{1}$, $C_{2}$, and $C_{3}$ that appear in
equations (\ref{Eq:VRGVF})-(\ref{Eq:VRGVGammaGP2}) are given by
\begin{eqnarray}
 \fl C_1 = \frac{2(v_\Delta/v_F)}{N \pi^3}\int^{\infty}_{-\infty}dx
\int^{2\pi}_{0}d\theta \frac{x^2-\cos^2\theta-(v_\Delta/v_F)^2
\sin^2\theta}{(x^2+\cos^2\theta+(v_\Delta/v_F)^2
\sin^2\theta)^2}\mathcal {G}(x,\theta),  \\
\fl C_2 = \frac{2(v_\Delta/v_F)}{N \pi^3}\int^{\infty}_{-\infty}dx
\int^{2\pi}_{0}d\theta   \frac{-x^2+\cos^2\theta-(v_\Delta/v_F)^2
\sin^2\theta}{(x^2+\cos^2\theta+(v_\Delta/v_F)^2
\sin^2\theta)^2}\mathcal{G}(x,\theta),  \\
\fl C_3 = \frac{2(v_\Delta/v_F)}{N\pi^3}\int^{\infty}_{-\infty}dx
\int^{2\pi}_{0}d\theta \frac{x^2+\cos^2\theta-(v_\Delta/v_F)^2
\sin^2\theta}{(x^2+\cos^2\theta+(v_\Delta/v_F)^2
\sin^2\theta)^2}\mathcal{G}(x,\theta),
\end{eqnarray}
with
\begin{eqnarray}
\fl \mathcal{G}^{-1} = \frac{x^2+\cos^2\theta}{\sqrt {x^2 +
\cos^2\theta + (v_\Delta/v_F)^2 \sin^2\theta}} + \frac{x^2 +
\sin^2\theta}{\sqrt{x^2 + \sin^2\theta + (v_\Delta/v_F)^2
\cos^2\theta}}.
\end{eqnarray}

\section{RG equation for specific heat\label{App:SpecificHeat}}

\begin{figure}[htbp]
\center
\includegraphics[width=2.82in]{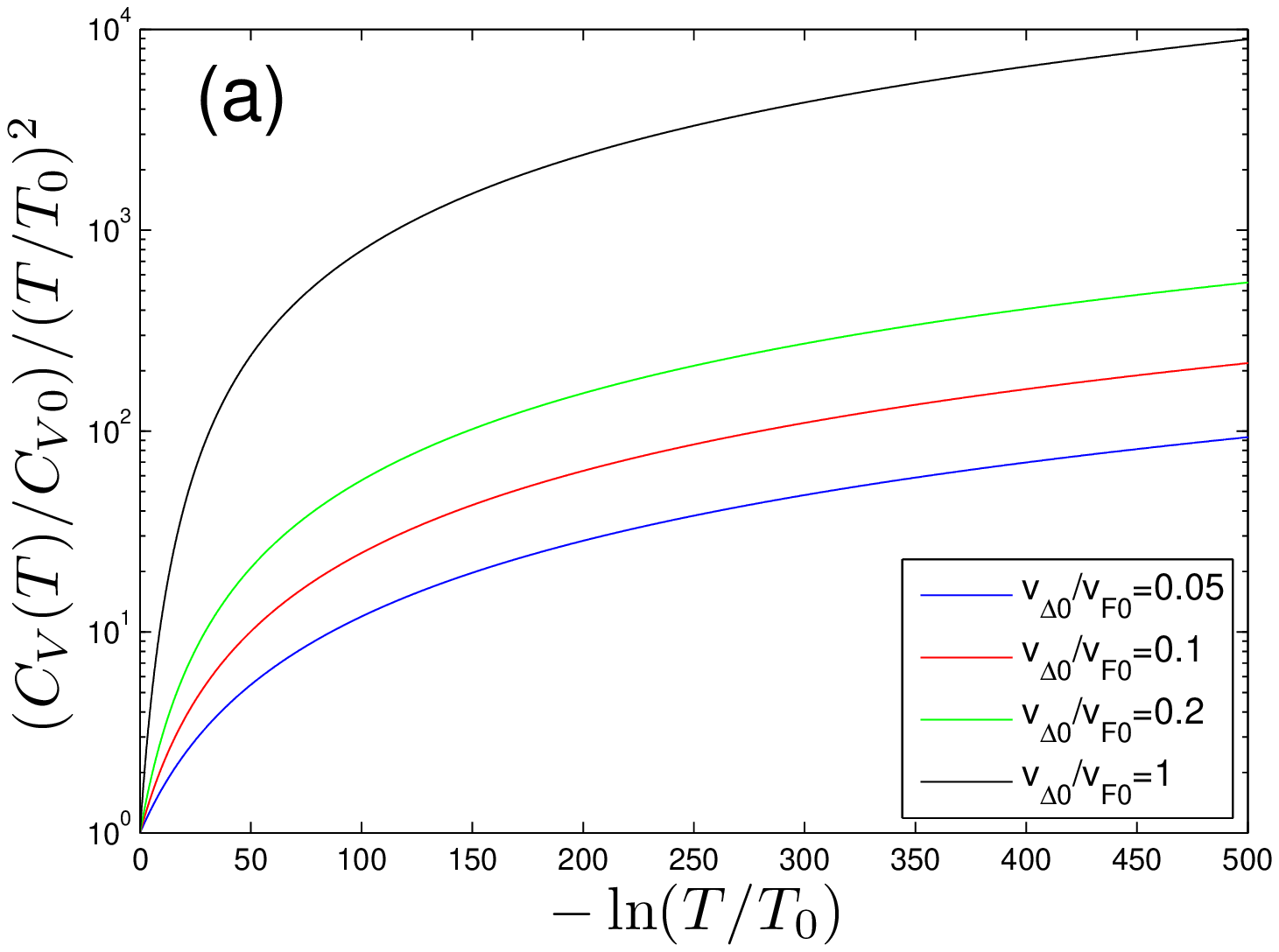}
\includegraphics[width=2.8in]{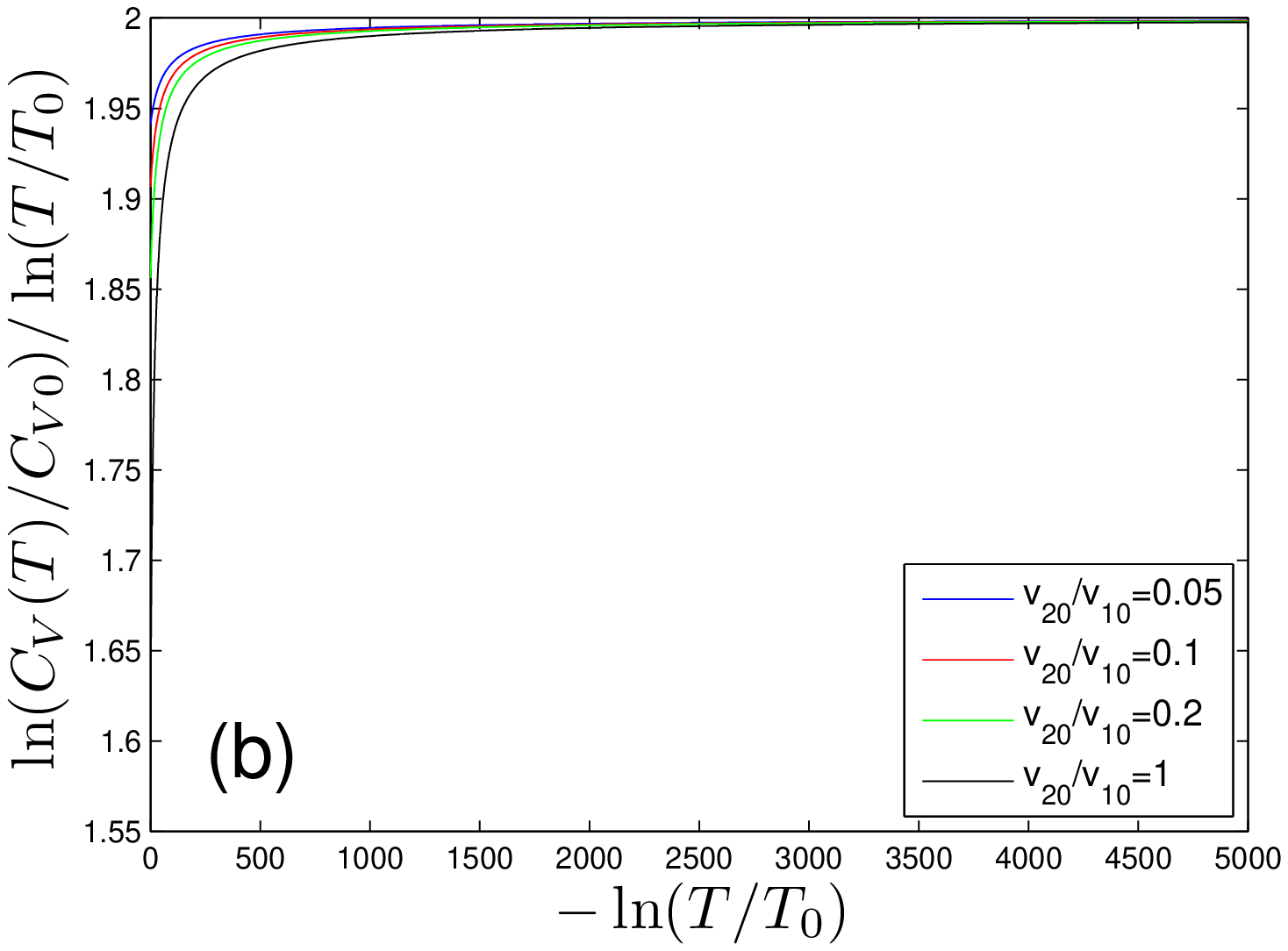}
\caption{Numerical results for $C_{V}(T)$ obtained from the RG
equation of specific heat given in \cite{Xu08} at different initial
values $v_{\Delta0}/v_{F0}=0.05, 0.1, 0.2, 1$. \label{Fig:CvClXu}}
\end{figure}

The free energy density $\mathcal{F} = T\ln\mathcal{Z}/V$ is
formally given by
\begin{eqnarray}
\mathcal{F} = \left(\xi_{\tau}\xi_{x}\xi_{y}\right)^{-1},
\end{eqnarray}
where $\xi_{\tau}\sim\frac{1}{T}$, $\xi_{x}=v_{F}\xi_{\tau}$, and
$\xi_{y} = v_{\Delta}\xi_{\tau}$. We can rewrite $\mathcal{F}$ as
\begin{eqnarray}
\mathcal{F}\sim\frac{1}{v_{F}v_{\Delta}}T^3,
\end{eqnarray}
which yields the following specific heat:
\begin{eqnarray}
C_{V} = -T\frac{\partial^2\mathcal{F}}{\partial T^2}\sim
\frac{1}{v_{F}v_{\Delta}}T^2.
\end{eqnarray}
It is then easy to get
\begin{eqnarray}
\frac{d\ln C_{V}}{d\ln T} = 2 +
\frac{d\ln\left(\frac{1}{v_{F}v_{\Delta}}\right)}{d\ln
T}\label{eqn:SpecificHeatMiddleA}.
\end{eqnarray}
At a given $T$, the corresponding momentum scale should be
determined by the larger component of the fermion velocities
\cite{Xu08}, i.e.,
\begin{eqnarray}
\tilde{p}=\frac{T}{\max(v_{F},v_{\Delta})},
\end{eqnarray}
which then gives rise to
\begin{eqnarray}
\frac{d\ln T}{d\ln\tilde{p}} = 1 + \frac{d\ln
\max(v_{F},v_{\Delta})}{d\ln\tilde{p}}.
\end{eqnarray}
The scaling equation for $C_{V}$ is converted to
\begin{eqnarray}
\frac{d \ln C_{V}}{d\ln T} &=& 2 + \frac{d\ln
\left(\frac{1}{v_{F}v_{\Delta}}\right)}{d\ln\tilde{p} \frac{d\ln
T}{d\ln\tilde{p}}} = 2 +\frac{d\ln\left(\frac{1}{v_{F}
v_{\Delta}}\right)}{d\ln\tilde{p} \left(1+\frac{d\ln
\max(v_{F},v_{\Delta})}{d\ln \tilde{p}}\right)}.
\end{eqnarray}
Since $d\ln\tilde{p}\sim -dl$, we find that
\begin{eqnarray}
\frac{d \ln C_{V}}{d\ln T} = 2 + \frac{\frac{d\ln
v_{F}}{dl}+\frac{d\ln v_{\Delta}}{dl}}{\left(1-\frac{d\ln
\max(v_{F},v_{\Delta})}{dl}\right)}.\label{eq:CVMiddle}
\end{eqnarray}
Substituting equations (\ref{Eq:VRGVF}) and (\ref{Eq:VRGVDelta})
into (\ref{eq:CVMiddle}) leads us to
\begin{eqnarray}
\frac{d \ln C_{V}}{d\ln T} = \left\{
\begin{array}{ll}
2 + \frac{2C_{1} - C_{2}-C_{3}-2C_{g}}{1-C_{1}+C_{2}+C_{g}},
&\mathrm{if}\quad v_{\Delta}/v_{F}<1,
\\
\\
2 + \frac{2C_{1} - C_{2}-C_{3}-2C_{g}}{1-C_{1}+C_{3}+C_{g}},
&\mathrm{if}\quad v_{\Delta}/v_{F}>1.
\end{array}\right.
\end{eqnarray}
The quantum fluctuation of nematic order drives the velocity ratio
$v_{\Delta}/v_{F}$ to monotonously decrease as one goes to lower and
lower energies. It is known that the bare ratio $v_{\Delta0}/v_{F0}
< 1$ in cuprate superconductors \cite{Orenstein00}, which make it
possible to simplify the RG equation to
\begin{eqnarray}
\frac{d \ln C_{V}}{d\ln T} = 2 + \frac{2C_{1} - C_{2} - C_{3} -
2C_{g}}{1 - C_{1} + C_{2} + C_{g}}.
\end{eqnarray}
In the clean limit, $C_{g} = 0$, so the equation becomes
\begin{eqnarray}
\frac{d \ln C_{V}}{d\ln T} = 2 + \frac{2C_{1} -
C_{2}-C_{3}}{1-C_{1}+C_{2}}.\label{Eq:RGEquationCvClean}
\end{eqnarray}
This equation is slightly different from the RG equation for
specific heat presented in \cite{Xu08}, where the equation is
\begin{eqnarray}
\frac{d \ln C_{V}}{d\ln T} = \frac{2+2C_{1} - C_{2} - C_{3}}{1 -
C_{1} + C_{2}}.\label{Eq:RGEquationCvXu}
\end{eqnarray}
The numerical solutions for equation (\ref{Eq:RGEquationCvXu}) are
shown in figure \ref{Fig:CvClXu}.

We notice that the specific heat obtained from equation
(\ref{Eq:RGEquationCvXu}) also satisfies $\lim_{T\rightarrow0}
C_{V}(T)/T^2 \rightarrow \infty$ and $\lim_{T \rightarrow
0}\ln\left(C_{V}(T)\right)/\ln(T) \rightarrow 2$, which indicates
that the specific heat is enhanced comparing to the case of
non-interacting nodal fermion system. However, as will be shown in
Appendix C, it cannot be expressed as a power law function. The
reason is that the three parameters $C_{1}$, $C_{2}$, and $C_{3}$
appearing in equation (\ref{Eq:RGEquationCvXu}) all vanish in the
lowest energy limit, rather than approaching certain finite values.

\section{Approximate analytical expressions of DOS and specific heat \label{App:ApproDOSCv}}

In order to show that the DOS and specific heat in clean limit do
not exhibit power law behaviors, we now derive their approximate
analytical expressions in the low-energy regime from the RG results.
The RG equation of DOS is given by
\begin{eqnarray}
\frac{d\ln\rho}{d\ln\omega} = \frac{1 + C_{1} - C_{2} - C_{3}}{1 -
C_{1} + C_{2}}
\end{eqnarray}
in the absence of disorders. In the lowest energy limit, we know
that the velocity ratio $v_{\Delta}/v_{F}\rightarrow 0$. The
parameters $C_{1}$, $C_{2}$, and $C_{3}$, being functions of
$v_{\Delta}/v_{F}$, all vanish \cite{Xu08} as
$v_{\Delta}/v_{F}\rightarrow 0$. Thus the RG equation can be
approximately written as
\begin{eqnarray}
\frac{d\ln\rho}{d\ln\omega} &\approx& 1 + 2C_{1} - 2C_{2} - C_{3}
\nonumber \\
&\approx& 1 - \frac{2}{N}\frac{v_{\Delta}}{v_{F}}a_{1} +
\frac{2}{N}\frac{v_{\Delta}}{v_{F}}a_{2}
-\left(\frac{8}{\pi^2}\ln\left(\frac{v_{F}}{v_{\Delta}}\right) -
a_{3}\right)\frac{1}{N}\frac{v_{\Delta}}{v_{F}} \nonumber \\
&\approx& 1 - \left(2a_{1}-2a_{2}-a_{3}\right)
\frac{1}{N}\frac{v_{\Delta}}{v_{F}} -\frac{8}{N\pi^2}
\ln\left(\frac{v_{F}}{v_{\Delta}}\right)\frac{v_{\Delta}}{v_{F}},
\label{Eq:DOSApproximateA}
\end{eqnarray}
where we have used three constants $a_{1}\approx0.426$,
$a_{2}\approx0.348$, and $a_{3}\approx0.96$ \cite{Huh08}.
Substituting the approximate low-energy expression of
$v_{\Delta}/v_{F}$ given in equation (\ref{Eq:VRatioApproximate})
into (\ref{Eq:DOSApproximateA}), we obtain
\begin{eqnarray}
\fl \frac{d\ln\rho}{d\ln\omega} &\sim& 1 -
\left(2a_{1}-2a_{2}-a_{3}\right)\frac{1}{N} \frac{\pi^2
N}{8}\frac{1}{l\ln(l)} - \frac{8}{N\pi^2}\ln\left(\frac{8}{\pi^2
N}l\ln(l)\right)\frac{\pi^2 N}{8}\frac{1}{l\ln(l)}\nonumber \\
\fl &\sim& 1 - a_{\rho}\frac{1}{l\ln(l)}
-\frac{1}{l}-\frac{\ln\ln(l)}{l\ln(l)}, \label{Eq:DOSApproximateB}
\end{eqnarray}
where $a_{\rho} = \left(2a_{1} - 2a_{2} -
a_{3}\right)\frac{\pi^2}{8} + \ln\left(\frac{8}{\pi^2 N}\right)$.
For $N=2$, $a_{\rho} \approx -1.896$. Using the relationship $\omega
= \omega_{0}e^{-l}$, we can solve equation
(\ref{Eq:DOSApproximateB}) and then obtain the following analytical
expression
\begin{eqnarray}
\rho(\omega) \sim \left(\frac{\omega}{\omega_{0}}\right)
\ln\left(\frac{\omega_{0}}{\omega}\right)
\left(\ln\ln\left(\frac{\omega_{0}}{\omega}\right)\right)^{a_{\rho}}
\exp \left[\frac{1}{2}\left(\ln\ln\ln
\left(\frac{\omega_{0}}{\omega}\right)\right)^{2}\right],
\end{eqnarray}
which is applicable for small $\omega$.

The RG equation for specific heat is shown in equation
(\ref{Eq:RGEquationCvClean}). At low energies, it can be
approximated as
\begin{eqnarray}
\frac{d\ln C_{V}}{d\ln T} &\sim& 2 + 2C_{1} - C_{2} - C_{3}\nonumber \\
&\sim& 2 - \frac{2}{N}\frac{v_{\Delta}}{v_{F}}a_{1} +
\frac{1}{N}\frac{v_{\Delta}}{v_{F}}a_{2} - \left(\frac{8}{\pi^2}
\ln\left(\frac{v_{F}}{v_{\Delta}}\right) -
a_{3}\right)\frac{1}{N}\frac{v_{\Delta}}{v_{F}} \nonumber \\
&\sim& 2 - a_{C}\frac{1}{l\ln(l)} - \frac{1}{l} -
\frac{\ln\ln(l)}{l\ln(l)},\label{Eq:CVApproximateB}
\end{eqnarray}
where $a_{C} = \left(2a_{1} - a_{2} - a_{3}\right)\frac{\pi^2}{8} +
\ln\left(\frac{8}{\pi^2 N}\right)$. At $N = 2$, $a_{C} \approx
-1.466$. Using the transformation $T = T_{0}e^{-l}$, we find that
the specific heat at low temperature can be well approximated by the
expression
\begin{eqnarray}
\fl C_{V}(T)\sim \left(\frac{T}{T_{0}}\right)^{2}
\ln\left(\frac{T_{0}}{T}\right)
\left(\ln\ln\left(\frac{T_{0}}{T}\right)\right)^{a_{C}} \exp
\left[\frac{1}{2}\left(\ln\ln\ln
\left(\frac{T_{0}}{T}\right)\right)^{2}\right].
\end{eqnarray}
By applying the same treatment, the RG equation of specific heat
presented in \cite{Xu08}, shown in equation
(\ref{Eq:RGEquationCvXu}), can be approximately expressed as
\begin{eqnarray}
\fl C_{V}(T)\sim \left(\frac{T}{T_{0}}\right)^{2}
\ln\left(\frac{T_{0}}{T}\right)
\left(\ln\ln\left(\frac{T_{0}}{T}\right)\right)^{a_{C}'} \exp
\left[\frac{1}{2}\left(\ln\ln\ln\left(\frac{T_{0}}{T}\right)\right)^{2}\right],
\label{Eq:CVApproximateXu}
\end{eqnarray}
where $a_{C}' = \left(4a_{1} - 3a_{2} - a_{3}\right) \frac{\pi^2}{8}
+ \ln\left(\frac{8}{\pi^2 N}\right)$. At $N=2$, $a_{C}' \approx
-1.273$. The above two functions do not exhibit power law dependence
on temperature.

\end{document}